\documentclass[final,onefignum,onetabnum]{siamltex1213}

\usepackage{blkarray}
\usepackage{amsmath}

\usepackage{dgleich-setup}
\usepackage{dgleich-math}
\renewcommand{\mat}{\boldsymbol}
\usepackage{epstopdf}
\usepackage{url}
\usepackage{tabularx}
\usepackage{paralist}
\usepackage{subfigure}
\usepackage{multicol}

\usepackage{tikz,nicefrac,amsmath,pifont}
\usepackage{tkz-graph}
\usetikzlibrary{arrows,backgrounds,patterns,matrix,shapes,fit,calc,shadows,plotmarks}
\usepackage{multirow}

\let\oldthebibliography=\thebibliography
\let\oldendthebibliography=\endthebibliography

\usepackage{natbib}
\renewcommand{\cite}{\citep}

\let\thebibliography=\oldthebibliography
\let\endthebibliography=\oldendthebibliography

  \providecolor{theblue}{RGB}{0,0,180}%
  \RequirePackage[colorlinks,pdfdisplaydoctitle
      ,citecolor=theblue
      ,linkcolor=theblue
      ,urlcolor=theblue
      ,hyperfootnotes=false,pagebackref]{hyperref}%
\renewcommand*{\backref}[1]{}%
\renewcommand*{\backrefalt}[4]{%
  \ifcase #1 %
    No citations.%
  \or
    Cited on page #2.%
  \else
    Cited on pages #2.%
  \fi
}%

\usepackage{todonotes}
\usetikzlibrary{arrows,backgrounds,patterns,matrix,shapes,fit,calc,shadows,plotmarks}
\makeatletter
\global\let\tikz@ensure@dollar@catcode=\relax
\makeatother

\usepackage{ushort}

\usepackage{epsdice}

\newcommand*{\vcenteredhbox}[1]{\begingroup
\setbox0=\hbox{#1}\parbox{\wd0}{\box0}\endgroup}

\usepackage{framed}
\usepackage{wrapfig}
\definecolor{shadecolor}{gray}{0.9}
\newcounter{AsideNumber}
\newcommand*{\aside}[2]{\stepcounter{AsideNumber}%
\label{#1}%
\begin{wrapfigure}{r}{0.4\linewidth}
\begin{shaded*}
\footnotesize%
\textsc{Aside~\theAsideNumber.~}\textit{#2}
\end{shaded*}
\end{wrapfigure}%
}

\usepackage{algorithm}
\usepackage[noend]{algpseudocode}
\usepackage[utf8]{inputenc}
\usepackage{listings}

\usepackage{bbding}
\usepackage{comment}
\newtheorem{problem}{Problem}
\newcommand*{\QED}{\hfill\ensuremath{\square}}

\graphicspath{{.}{./FIG/}{./figures/}}

\newcommand{\mKk}{\mK \kron \overset{k \text{ terms}}{\ldots} \kron \mK}

\newcommand{\mMt}{\underline{\mM}}

\usepackage[frozencache=true]{minted}

\newcommand{\ERfull}{Erd\H{o}s-R\'{e}nyi\xspace}

\newenvironment{solution}{\smallskip \par \emph{Solution.}}{\hfill\checkmark\bigskip}

\title{Coin-flipping, ball-dropping, and grass-hopping for generating random graphs from matrices of edge probabilities\thanks{Funding: Partially supported by the NSF funded NSF CAREER award CCF-1149756, NSF Big data IIS-1546488, NSF Center for Science of Information STC, CCF-093937, DARPA SIMPLEX and the Sloan Foundation.}}
\author{Arjun~S.~Ramani\thanks{West Lafayette Jr/Sr High School, 1105 North Grant Street, West Lafayette, IN 47906, USA (\email{aramani@purdue.edu}).} \and Nicole Eikmeier\thanks{Department of Mathematics, Purdue University, \email{eikmeier@purdue.edu}}  \and David~F.~Gleich\thanks{Department of Computer Science, Purdue University, 305 North University Avenue, West Lafayette, IN 47907, USA (\email{dgleich@purdue.edu}).}}

\date{}
\begin{document}
\maketitle


\begin{abstract}
Common models for random graphs, such as \ERfull and Kronecker graphs,
correspond to generating random adjacency matrices where each entry is non-zero
based on a large matrix of probabilities.
Generating an instance of a random graph based
on these models is easy,
although inefficient, by flipping biased coins (i.e.~sampling 
binomial random variables) for each possible edge. This process
is inefficient because most large graph models correspond to sparse
graphs where the vast majority of coin flips will result in no edges. 
We describe some not-entirely-well-known, but 
not-entirely-unknown, techniques that will enable us to sample a 
graph by finding only the coin flips that will produce edges.
Our analogies for these procedures are ball-dropping, which is easier
to implement, but may need extra work due to duplicate edges, and 
grass-hopping, which results in no duplicated work or extra edges. 
Grass-hopping does this using geometric random variables.
In order to use this idea on complex 
probability matrices such as those in Kronecker graphs, we decompose
the problem into three steps, each of which are independently useful
computational primitives: (i) enumerating non-decreasing sequences, 
(ii) unranking multiset permutations, and (iii) decoding and encoding
z-curve and Morton codes and permutations. The third step is the result of
a new connection between repeated Kronecker product operations
and Morton codes. Throughout, 
we draw connections to ideas underlying applied math and computer science
including coupon collector problems. 
\end{abstract}

\makeatletter
\newenvironment{audience}{\begin{@abssec}{Audience}}{\end{@abssec}}
\makeatother

\begin{audience} 
This paper is designed, primarily, for undergraduates who have some experience with mathematics classes at the level of introduction to probability, discrete math, calculus, and programming, as well as for educators teaching these classes. We try and provide pointers to additional background material where relevant -- and we also provide links to various courses throughout applied math and computer science to facilitate using this material in multiple places. Our program codes are available through the github repository: \url{https://github.com/dgleich/grass-hopping-graphs}.
\end{audience}

\section{Introduction \textit{\&} Motivation for Fast Random Graph Generation}


The utility of a random graph is akin to the utility of a random number. 
Random numbers, random variables, and the framework of statistical
hypothesis testing provide a convenient way to study and assess whether an apparent signal or pattern extracted from noisy data is likely to be real or the result of chance.  In this setting, a random variable 
models the null-hypothesis where the effect is due to chance alone.
\begin{figure}
\vcenteredhbox{\begin{tabularx}{0.65\linewidth}{lXXXXXXl}
\toprule
& \Large\epsdice{1} & \Large\epsdice{2} & \Large\epsdice{3} & \Large\epsdice{4} & \Large\epsdice{5} & \Large\epsdice{6} &\\
\midrule
Measured & 30 & 32 & 33 & 31 & 29 & 25 \\ \addlinespace 
Expected & 30 & 30 & 30 & 30 & 30 & 30 \\
\midrule
Statistic & 0 & 2/15 & 3/10 & 1/30 & 1/30 & 5/6 & 4/3 \\
\bottomrule 
\end{tabularx}}%
\vcenteredhbox{\includegraphics[width=0.35\linewidth]{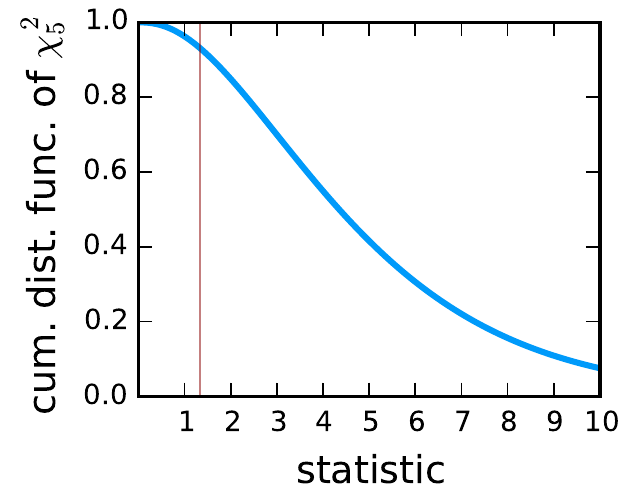}}


\caption{An example of statistical hypothesis testing in the classic scenario with a closed form solution. Here, the closed form $\chi^2$ test works to determine if a dice is ``fair''. In this case, we conclude there is a 93\% chance the dice is fair. For random graph models, we often have no analytic way to determine the quantities akin to the \emph{expected} number or the values from the cumulative distribution function. Thus, we need to \emph{quickly} generate a large number of random graphs to generate simulation based estimates instead.}
\label{fig:dice}
\end{figure} 
By way of example, testing if a standard six-sided dice is ``fair'' involves a random variable with a $\chi^2$ distribution. The way this test works is illustrated in Figure~\ref{fig:dice}. We compare the number of times the dice rolls each number with the expected number of times under the \emph{fair} hypothesis. For each outcome, such as \epsdice{3}, we compute $(\text{Measured} - \text{Expected})^2/\text{Expected}$. Then we sum these over all outcomes. The resulting number is a \emph{statistic} called Pearson's cumulative test statistic. (In the figure, this is $4/3$.) The test statistic is also a random variable because if we rolled the dice another set of times, we could get a different value. A random variable with a $\chi^2$ distribution is the behavior of this test statistic \emph{when the dice is exactly fair}.

\aside{}{This type of analysis and methodology is called statistical hypothesis testing and can be extended to more complex scenarios. Estimating these distributions from test statistics results in studies which are fascinating in their own right. } The famous $p$-value is the probability that a fair die would give rise to a test statistic value \emph{greater} than the \emph{computed} test statistics (which is $4/3$). In other words, we look at the probability that a value from a $\chi^2$ distribution is bigger than 4/3. The resulting probability ($p = 0.931$) provides a convenient estimate of how likely the dice is to be fair. In this case, we computed the probability using an analytic description of the $\chi^2$ distribution (strictly speaking, this test involves the $\chi^2$ distribution with $5$ degrees of freedom). The probability should be large if the dice is fair and small if it is unlikely to be fair. Often, the probability $p=0.05$ is arbitrarily chosen as a cut-point to distinguish the two cases ($p \ge 0.05$ means fair and $p < 0.05$ means maybe not fair), but this value should be used judiciously. 

Statistical hypothesis testing on graphs is largely the same~\cite{Moreno-2013-hypothesis,Koyuturk-2007-significance,Koyuturk-2006-significance,milo}. We use a random graph model as the null hypothesis. This is akin to the \emph{fair} hypothesis used in the example above. Then we measure something on a real-world graph and compare what we would have expected to get on a set of random graphs. Consider, for instance, the study by \citet{milo} on the presence of \emph{motifs} in networks. This study defines network motifs as patterns of connectivity that are found more frequently in a network than what would be expected in a random network. In social networks such as Facebook, a common motif is a triangle. If you have two friends, then it is much more likely that your friends are also friends, which forms a triangle structure in the graph. This property does not exist in many random models of networks. 
\aside{triangle-models}{In fact, creating random graph models with the property that they have a non-trivial number of triangles is an active area of study! Recent work in this vein includes \citet{Kolda-2014-BTER,Newman-2009-random-graphs}.} 
One critical difference from traditional statistics, however, is that we often lack closed-form representations for the probability of many events in random graph models. There was \emph{no known expression} for the number of motifs that Milo should expect unlike the case of the dice and the $\chi^2$ distribution. In fact, in the studies above, only \cite{Koyuturk-2006-significance,Koyuturk-2007-significance} had closed form solutions. 

Thus, we turn to empirical simulations. One of the easiest and most common methods to do these studies (such as \citet{milo}) leverages the computer to generate hundreds or hundreds of thousands of instances of a random graph and compute their properties. In comparison with the dice example, if we had access to a dice that was guaranteed to be fair, we could have generated an empirical distribution for Pearson's cumulative test statistic by tossing the fair dice a few thousand times. Then, an estimate of the probability that the statistic is larger could be computed just by checking the number of instances where the sample was larger. 
\aside{permutation-tests}{Indeed, \citet{Fisher-1936-racial} mentions this ``trivial'' toss-dice-and-check idea as the basis of all rigorous work in the area of hypothesis testing. More recently, as the power of computation has grown, others including \citet{Ernst-2004-permutation} have suggested renewed focus on these empirical and computational tests rather than analytic results.} To do this, we would need a \emph{fast} way of tossing dice! For random graphs, it almost goes without saying that we also need this random graph generation, which is also called random graph sampling, to be as fast and efficient as possible.

Statistical hypothesis testing is not the only use for random graphs. There are a variety of other scenarios that also require fast sampling such as benchmarking high performance computers~\cite{Murphy-2010-graph500} and synthetic data to evaluate algorithms~\cite{Kloumann-2016-block-models}.  What distinguishes the statistical hypothesis testing scenario is that it is important to \emph{exactly sample the true random graph model}. As we shall see, there are a variety of efficient methods that can slightly bias these results or may take additional effort to compute.

In this paper, we will explain and illustrate how to generate random graphs efficiently when the random graph model is described by a matrix of probabilities for each edge. This is admittedly a special case, but it handles some of the most widely used random graph distributions:
\begin{itemize}[\quad-]
\item \emph{the \ERfull model} (\S\ref{sec:er}),
\item \emph{the Kronecker model} (\S\ref{sec:kron}),
\item \emph{the Chung-Lu model} (\S\ref{sec:chung-lu}), and
\item \emph{the stochastic block model} (\S\ref{sec:sbm}).
\end{itemize}
We will pay special attention to some of the structures present in the matrices of probabilities of stochastic Kronecker graphs and how to take advantage of them to make the process go faster. Along the way, we'll see examples of a number of classic discrete structures and topics:
\begin{itemize}[\quad-]
\item \emph{binomial and geometric random variables},
\item \emph{the coupon collector problem},
\item \emph{enumerating non-decreasing sequences},
\item \emph{unranking multiset permutations}, and
\item \emph{Morton and Z-curves}.
\end{itemize}
Throughout, we are going to explain these concepts and ideas both mathematically and programmatically. We include runnable Python code in the manuscript to make it easy to follow along, see \url{https://github.com/dgleich/grass-hopping-graphs/} for our code.

\section{Random Graph Models and Matrices of Probabilities} 

A graph consists of a set of vertices $V$ and a set of edges $E$, where each edge $e \in E$ encodes relationships between two vertices, which is often written $e = (u,v)$ for $u, v \in V$. We consider the more general setting of directed graphs in this paper, although we describe how all of the ideas specialize to undirected graphs in Section~\ref{sec:undirected}.  A random graph consists of a set of random edges between a fixed number of vertices. 
\aside{general-models}{There are more general definitions of random graphs that do not impose a fixed number of vertices such as preferential attachment~\cite{barabasi1999-scaling}, but this simple setting will serve our purposes.}
How should these random edges be chosen? That turns out to depend on the particular \emph{random graph model}. 

The models we consider generate a \emph{random adjacency matrix} where each entry $A_{ij}$ in the adjacency matrix is $0$ or $1$ with probability $P_{ij}$ for some given matrix of probabilities $P$. Let's dive into these details to understand exactly what this means.

\subsection{The adjacency matrix} 
\label{sec:adj}
The adjacency matrix encodes the information of the nodes and edges into a matrix. For instance: 
\begin{center}
\vcenteredhbox{if the graph is }
\vcenteredhbox{
\sffamily

\begin{tikzpicture}[scale=0.28]
    \renewcommand*{\VertexInterMinSize}{12pt}
  \renewcommand*{\VertexSmallMinSize}{0pt}
\Vertex[x=0,y=3]{A}
\Vertex[x=2,y=0]{B}
\Vertex[x=3,y=6]{C}
\Vertex[x=7,y=5.5]{D}
\Vertex[x=6.5,y=.25]{E}
\Vertex[x=10,y=4]{F}
\Vertex[x=9.5,y=2]{G}
\SetUpEdge[style={post}]
\Edge(A)(C)
\Edge(C)(B)
\Edge(F)(D)
\Edge(D)(E)
\Edge(E)(C)
\Edge(D)(E)
\Edge(B)(D)
\Edge(D)(G)
\Edge(B)(E)
\SetUpEdge[style={post,bend right}]
\Edge(C)(D)
\Edge(D)(C)
\Edge(A)(B)
\Edge(B)(A)
\end{tikzpicture}
}
\vcenteredhbox{, the adjacency matrix is}
\vcenteredhbox{\scalebox{0.9}{\arraycolsep=1.4pt\fontsize{8}{8.25}\selectfont $\begin{blockarray}{cc@{\,}c@{\,}c@{\,}c@{\,}c@{\,}c@{\,}c}
& A & B & C & D & E & F & G \\
\begin{block}{c[c@{\,}c@{\,}c@{\,}c@{\,}c@{\,}c@{\,}c]}
A & 0 & 1 & 1 & 0 & 0 & 0 & 0 \\
B & 1 & 0 & 0 & 1 & 1 & 0 & 0 \\
C & 0 & 1 & 0 & 1 & 0 & 0 & 0 \\
D & 0 & 0 & 1 & 0 & 1 & 0 & 1 \\
E & 0 & 0 & 1 & 0 & 0 & 0 & 0 \\
F & 0 & 0 & 0 & 1 & 0 & 0 & 0  \\
G & 0 & 0 & 0 & 0 & 0 & 0 & 0  \\
\end{block}
\end{blockarray}$}.}
\end{center}
Formally, the adjacency matrix is created by assigning each vertex $v \in V$ a unique number between $1$ and $|V|$ often called an index. Then each edge $e = (u,v) \in E$ produces an entry of $1$ in the coordinate of the matrix that results in mapping $u$ and $v$ to their indices. That is, if $i$ and $j$ are the indices of $u$ and $v$, then entry $i,j$ of the matrix has value $1$. All other entries of the matrix are $0$. In the case above, we mapped $A$ to index $1$, $B$ to index $2$ and so on. 

\subsection{A random adjacency matrix as a random graph}
\aside{matrix-as-graph}{In fact, there are many aspects of sparse matrix computations that are most easily understood by viewing a sparse matrix as a graph and computing a graph algorithm such as breadth-first search~\cite{davis2006-book}.}
Note that this process can go the other way as well. If we have any $n$-by-$n$ matrix where each entry is $0$ or $1$, then we can interpret that matrix as the adjacency matrix of a graph! What we are going to do is (i) generate a random matrix with each entry being $0$ or $1$ and (ii) interpret that matrix as a set of random edges to give us the random graph. 

At this point, we need to mention a distinction between \emph{directed} and \emph{undirected graphs}. By convention, an undirected graph has a symmetric adjacency matrix where $A_{ij} = 1$ and $A_{ji} = 1$ for each undirected edge. Our focus will be on generating random \emph{non-symmetric} matrices $\mA$. These techniques will still apply to generating symmetric graphs, however. For instance, we can interpret a non-symmetric $\mA$ as an undirected graph by only considering entries $A_{ij}$ where $i < j$ (or equivalently, $A_{ij}$ where $i > j$) and then symmetrizing the matrix given one of these triangular regions. (In Matlab, this would be: \verb#T = triu(A,1); G = T+T'#; in Python, it would be \verb#np.triu(A,1); G=T+T.T#) We return to this point in Section~\ref{sec:undirected}. 

\subsection{The \ERfull\ model} 
\label{sec:er}

Perhaps the first idea that comes to mind at this point is tossing a coin to determine if an entry in the adjacency matrix should be $0$ or $1$. While a simple 50\%-50\% coin toss is suitable for randomly picking between the two, there is no reason we need $1$'s (the edges) to occur with the same probability as $0$'s (the non-edges). Let $p$ be the probability that we generate a $1$ (an edge) and let 
\begin{equation}
A_{ij} = \begin{cases} 1 & \text{ with probability $p$} \\ 0 & \text{ with probability $1-p$} \end{cases} \text{ for each } i,j \text{ in an $n$-by-$n$ matrix}. 
\end{equation}
\aside{note-er}{The \ERfull graph as we consider it was actually proposed by Edgar Gilbert~\protect\cite{Gilbert-1959-random} at the same time at Erd\H{o}s and R\'{e}nyi's famous paper~\protect\cite{Erdos-1959-random}. Erd\H{o}s and R\'{e}nyi proposed a slight variation that fixed the number of edges and placed them uniformly at random.}
This model is called the \ERfull model. It is one of the simplest types of random graphs and has $n$ nodes where any 2 nodes have probability $p$ of being connected with a directed edge. \ERfull graphs are traditionally constructed by considering each edge separately and using the method of coin-flipping, which will be discussed in Section~\ref{sec:coin-flipping}.

\subsection{Random graph models as matrices of probabilities} 
\label{sec:random-adj}
Now, you might be wondering why we use only a \emph{single} probability $p$ for each edge in the \ERfull model. The random graph models we study here have an entry $P_{ij}$ for each $i,j$, which is a more general setting than the \ERfull construction:
\begin{equation}
A_{ij} = \begin{cases} 1 & \text{ with probability $P_{ij}$} \\ 0 & \text{ with probability $1-P_{ij}$} \end{cases} \text{ for each } i,j \text{ in an $n$-by-$n$ matrix}. 
\end{equation}
Thus, if we set $P_{ij} = p$, then this more general model corresponds to the \ERfull model. Now, of course, this begs two questions. How do we choose $P_{ij}$? How do we generate a random matrix $\mA$? We will describe three common random graph models that consist of a choice of $P_{ij}$: the Kronecker model, the Chung-Lu model, and the stochastic block models.  We will also describe the coin-flipping method of sampling a random graph at this point. However, in subsequent sections, we will show how to sample each of these models more efficiently than the coin-flipping method. 

\subsection{The coin-flipping method for sampling a random graph}
\label{sec:coin-flipping}
Given the matrix of probabilities $\mP$, the easiest way to generate a random adjacency matrix where $A_{ij}$ is $1$ with probability $P_{ij}$ is to explicitly simulate biased coin flips (which are also Bernoulli random variables). 
\aside{pseudorandom}{Computers typically use pseudo-random number generators which have their own fascinating field~\cite{Gentle-2003-random-numbers,Knuth-1997-art-vol-2}.} 
A pragmatic way to do this is to use a random number generator that produces a uniform distribution of values between $0$ and $1$. This is pragmatic because most computer languages include such a routine and make sure it is efficient. Given such a random value $\rho$, we set $A_{ij}$ to $1$ if $\rho \le P_{ij}$ (which happens with probability $P_{ij}$). An example of doing this is show in Listing~\ref{code:coin-flip}. Note that the indices in Python range from $0$ to $n-1$ rather than from $1$ to $n$ as is common in mathematical descriptions of matrices. 

\begin{listing}
\caption{A simple code to generate a random graph by coin flipping}
\label{code:coin-flip}
\begin{minted}[fontsize=\footnotesize]{python}
import random       # after this, random.random() gives a uniform [0,1] value
import numpy as np  # numpy is the Python matrix package
"""
Generate a random graph by coin flipping. The input is a square matrix P with entries 
between 0 and 1. The result is a 0 or 1 adjacency matrix for the random graph. 

Example: 
  coinflip(0.25*np.ones((8,8))) # generate a random Erdos-Renyi graph
"""
def coinflip(P):
  n = P.shape[0]
  assert(n == P.shape[1]) # make sure we have a square input
  A = np.zeros_like(P)    # create an empty adjacency matrix
  for j in range(n):
    for i in range(n):    # fill in each entry as 1 with prob P[i,j]
      A[i,j] = random.random() <= P[i,j]  
  return A      
\end{minted} 
\end{listing}

\subsection{The Kronecker model} 
\label{sec:kron}
The Kronecker random graph model results in a non-uniform but highly structured probability matrix $\mP$. It begins with a small initiator matrix $\mK$ with $n$ nodes and then enlarges it to a bigger matrix of probabilities $\mP$ by taking successive Kronecker products~\cite{Chakrabarti-2004-rmat,leskovec2005,leskovec2010}. For an example suppose $\mK$ is a $2 \times 2$ initiator matrix
\begin{equation}
\mK=
\begin{bmatrix}
	a&b\\
    c&d\\
\end{bmatrix},
\end{equation}
where each $a, b, c, d$ is a probability. 
The Kronecker product of $\mK$ with itself is given by
\[
\mK \otimes \mK = 
\begin{bmatrix}
a\cdot \mK &b\cdot \mK\\ c\cdot \mK &d\cdot \mK
\end{bmatrix}=
\begin{bmatrix}
aa&ab&ba&bb \\ ac&ad&bc&bd \\ ca&cb&da&db \\ cc&cd&dc&dd
\end{bmatrix}.
\]
Finally, the $k$\textsuperscript{th} Kronecker product of $\mK$ is just 
\begin{equation} \underbrace{\mK \otimes \mK \otimes \ldots \otimes \mK}_{k\, \text{times}},
\end{equation}
which is a $2^k$-by-$2^k$ matrix. As a concrete example, let $\mK = \sbmat{0.99 & 0.5 \\ 0.5 & 0.2}$, then 
\begin{equation}
\label{eq:kron-example}
\mK \kron \mK = \sbmat{
 0.9801 &  0.495 &  0.495 & 0.25\\
 0.495 &  0.198 & 0.25 &  0.1  \\
 0.495 &  0.25 &  0.198 &  0.1 \\
 0.25  &  0.1 &   0.1 &   0.04}, 
 \quad
 \mK \kron \mK \kron \mK  = \sbmat{
 0.97 &0.49 &0.49 &0.25 &0.49 &0.25 &0.25 &0.13 \\ 
0.49 &0.20 &0.25 &0.10 &0.25 &0.10 &0.13 &0.05 \\ 
0.49 &0.25 &0.20 &0.10 &0.25 &0.13 &0.10 &0.05 \\ 
0.25 &0.10 &0.10 &0.04 &0.13 &0.05 &0.05 &0.02 \\ 
0.49 &0.25 &0.25 &0.13 &0.20 &0.10 &0.10 &0.05 \\ 
0.25 &0.10 &0.13 &0.05 &0.10 &0.04 &0.05 &0.02 \\ 
0.25 &0.13 &0.10 &0.05 &0.10 &0.05 &0.04 &0.02 \\ 
0.13 &0.05 &0.05 &0.02 &0.05 &0.02 &0.02 &0.01 \\ 
 }.
 \end{equation}
We use these $k$\textsuperscript{th} Kronecker products as the matrix of probabilities for the random graph. This gives us (in general) an $n^k$-by-$n^k$ matrix of probabilities $\mP$ for an $n^k$-node random graph. 
It quickly becomes tedious to write out these matrices by hand. Surprisingly, there \emph{is structure} inside of this matrix of repeated Kronecker products, and we will return to study and exploit its patterns in Section~\ref{sec:fastkron}. 

There are a variety of motivations for repeated Kronecker products as a graph model. On the practical side, they are extremely parsimonious and require only the entries of a small $n$-by-$n$ matrix, where $n$ would be between $2$ and $5$. Second, they can generate a variety of graphs of different sizes by adjusting $k$. Third, the graphs they produce have a number of highly skewed properties~\cite{Seshadhri-2013-kronecker}. These reasons make the Kronecker model a useful synthetic network model for various real-world performance studies~\cite{Murphy-2010-graph500}. On the statistical side, Kronecker models have some of the same properties as real-world networks~\cite{leskovec2010}. As such, they provide non-trivial null models. 


\subsection{The Chung-Lu model}
\label{sec:chung-lu} 
Recall that the degree, $d_u$, of a vertex $u$ is just the number of edges leaving $u$. For example, in the graph from Section~\ref{sec:adj}, the degrees of the nodes in alphabetical order are $(2,3,2,3,1,1,0)$. Many important and interesting features of a graph are deeply connected to the degrees of the vertices~\cite{Adamic-2001-search,litvak2006-pagerank-indegree}. This fact is the motivation for the Chung-Lu random graph model. We wish to have a random graph with vertices of roughly the same degree as a network we are studying to understand if the properties we observe (in the real network) are due to the degrees or the network structure. This is almost exactly what \citet{milo} did when they wanted to understand if a motif pattern was significant. 

In the Chung-Lu model, we need, as input, the desired degree of each vertex in the resulting random graph. For example, say we want to generate an $n$-by-$n$ graph where vertex 1 has degree $d_1$, vertex 2 has degree $d_2$ and so on. 
Specifically, suppose we want an $8$ vertex network with $1$ node of degree 4, 
$1$ node of degree $3$, $3$ nodes of degree $2$, and $3$ nodes of degree $1$. The
corresponding matrix is: 
\[ \mP = \sbmat{
1.00 &0.75 &0.50 &0.50 &0.50 &0.25 &0.25 &0.25 \\ 
0.75 &0.56 &0.38 &0.38 &0.38 &0.19 &0.19 &0.19 \\ 
0.50 &0.38 &0.25 &0.25 &0.25 &0.13 &0.13 &0.13 \\ 
0.50 &0.38 &0.25 &0.25 &0.25 &0.13 &0.13 &0.13 \\ 
0.50 &0.38 &0.25 &0.25 &0.25 &0.13 &0.13 &0.13 \\ 
0.25 &0.19 &0.13 &0.13 &0.13 &0.06 &0.06 &0.06 \\ 
0.25 &0.19 &0.13 &0.13 &0.13 &0.06 &0.06 &0.06 \\ 
0.25 &0.19 &0.13 &0.13 &0.13 &0.06 &0.06 &0.06 \\ }.
\]
This matrix results from setting
\begin{equation}
\label{eqn:chung-lu}
P_{ij} = \frac{d_i d_j}{\sum_k d_k}.
\end{equation} 

To understand why this is a good choice, let's briefly consider the \emph{expected degree} of vertex $i$. In the adjacency matrix, we can compute the degree by taking a sum of all entries in a row. Here, we have, in expectation: 
\begin{equation}
\label{eqn:chungexp}
\textstyle \text{E}[ \sum_{j} A_{ij}] =  \sum_{j} \text{E}[A_{ij}] = \sum_{j} P_{ij} = \sum_{j} \frac{d_i d_j}{\sum_k d_k} = d_i. 
\end{equation}
This analysis shows that, in expectation, this choice of probabilities result in a random graph with the correct degree distribution. 




\subsection{The stochastic block model}
\label{sec:stochastic}
\label{sec:sbm}
Another feature of real-world networks is that they have \emph{communities}~\cite{Flake-2000-communities,Newman2004-community}. A community is a group of vertices that are more tightly interconnected than they are connected to the rest of the graph. Pictorially this results in a graph such as: 
\begin{center}
\includegraphics[width=2in]{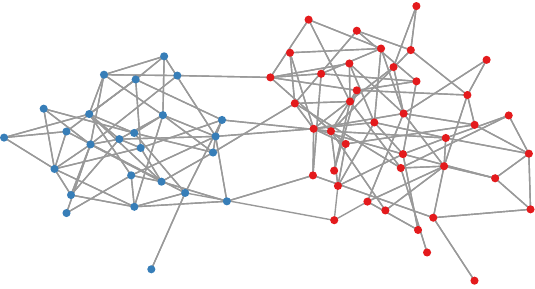}
\end{center}
where there are two communities: blue and red.
The stochastic block model is designed to mirror this structure in a random graph. Suppose we want a random graph with two communities with $n_1$ and $n_2$ nodes respectively. We want there to be high probability, $p$, of edges within a community, and a low probability of edges between the communities, $q < p$. 
Formally, this corresponds to a probability matrix: 
\begin{equation}
P_{ij} = \begin{cases} p & \text{if $i,j$ are in the same community} \\ q & \text{if $i,j$ are in different communities} \end{cases} \text{, \,  where $p> q$}. 
\end{equation}
The reason this is called the stochastic block model is that it can be written with a set of \emph{block} matrices. Consider $n_1 = 3$, $n_2 = 5$, $p = 0.7$, and $q = 0.1$:  
\begin{equation}
\mP =
\sbox0{$\begin{matrix}1&2\\0&1             \end{matrix}$}
\left[
\begin{array}{c|c}
\vphantom{\usebox{0}}\makebox[\wd0]{\large $p$}&\makebox[\wd0]{\large $q$} \\
\hline
\vphantom{\usebox{0}}\makebox[\wd0]{\large $q$}&\makebox[\wd0]{\large $p$}
\end{array}
\right]
= \arraycolsep=1.4pt\fontsize{8}{8.5}\selectfont
\left[ \begin{array}{ccc|ccccc}
0.7 & 0.7 & 0.7 & 0.1 & 0.1 & 0.1 & 0.1 & 0.1 \\
0.7 & 0.7 & 0.7 & 0.1 & 0.1 & 0.1 & 0.1 & 0.1 \\
0.7 & 0.7 & 0.7 & 0.1 & 0.1 & 0.1 & 0.1 & 0.1 \\
\hline
0.1 & 0.1 & 0.1 & 0.7 & 0.7 & 0.7 & 0.7 & 0.7 \\ 
0.1 & 0.1 & 0.1 & 0.7 & 0.7 & 0.7 & 0.7 & 0.7 \\ 
0.1 & 0.1 & 0.1 & 0.7 & 0.7 & 0.7 & 0.7 & 0.7 \\ 
0.1 & 0.1 & 0.1 & 0.7 & 0.7 & 0.7 & 0.7 & 0.7 \\ 
0.1 & 0.1 & 0.1 & 0.7 & 0.7 & 0.7 & 0.7 & 0.7 
\end{array}
\right].
\end{equation}

Notice that each of the blocks is just an \ERfull matrix. The model can be extended to an arbitrary number of subsets and the values of $p$ and $q$ can be varied between different subsets. In the more general case, let $Q_{rs}$ denote the probability of an edge between a node in the $r$th block and the $s$th block. Then the block adjacency matrix will look like:
%
\def\Big#1{\makebox(0,0){\huge#1}}
\[
\sbox0{$\begin{matrix}1&2\\0&1 \end{matrix}$}
  \begin{blockarray}{ccccc}
    &  n_1 & n_2 &  & n_k \\        
    \begin{block}{c[c|c|c|c]}       
n_1 & \vphantom{\usebox{0}}\makebox[\wd0]{\large $Q_{11}$}&\makebox[\wd0]{\large $Q_{12}$}  & \makebox[\wd0]{\large $\cdots$}& \makebox[\wd0]{\large $Q_{1k}$} \\\cline{2-5}      
n_2 & \vphantom{\usebox{0}}\makebox[\wd0]{\large $Q_{21}$} & \makebox[\wd0]{\large $Q_{22}$} & & \\ \cline{2-5}    
    & \vphantom{\usebox{0}}\makebox[\wd0]{\large $\vdots$} & \makebox[\wd0]{\large $\vdots$} & \makebox[\wd0]{\large $\ddots$}&\\ \cline{2-5}
n_k & \vphantom{\usebox{0}}\makebox[\wd0]{\large $Q_{k1}$} & \makebox[\wd0]{\large $Q_{k2}$} & & \makebox[\wd0]{\large $Q_{kk}$} \\        
  \end{block}
  \end{blockarray}.
\]

\subsection{Undirected graphs}
\label{sec:undirected}
The focus of our paper is on generating random graphs. To do this, we generate a random binary and non-symmetric adjacency matrix $\mA$. However, many studies on real-world data start with undirected graphs with symmetric matrices of probabilities $\mP$. Thus, we would like the ability to generate undirected graphs! We have three ways to get an undirected graph (and it's symmetric adjacency matrix) from these non-symmetric adjacency matrices. 

The first way has already been mentioned: take the upper or lower triangular piece of the non-symmetric matrix $\mA$ and just explicitly symmetrize that. (See Section~\ref{sec:random-adj}.) 

The second way builds upon the first way. Recall that in many scenarios discussed in the introduction, we are interested in generating a large number of random graphs from the same distribution. In this case, we note that a non-symmetric adjacency matrix $\mA$ that results from a symmetric matrix $\mP$ will give us \emph{two} samples of a random graph! One from the lower-triangular entries and one from the upper-triangular entries. \emph{This is the easiest and most pragmatic way to proceed if you are generating multiple samples as all of the code from this paper directly applies.} 

The third way proceeds by only generating edges in the upper-triangular region itself. For instance, in Listing~\ref{code:coin-flip}, we could easily restrict the \verb#for# loops to $i < j$, and set $A_{ji} = 1$ whenever we set $A_{ij} = 1$. This same strategy, however, will become much more complicated when we look at some of the accelerated generation schemes in subsequent sections. It is possible and it is an exercise worth exploring for those interested, but we deemed the added layer of complexity too high for this particular manuscript. Besides, the second method above is likely to be more useful in scenarios where multiple samples are needed and is also very easy to implement.


\section{Efficiently Generating Edges for \ERfull Graphs: Ball-dropping and Grass-hopping}

Recall that in an \ERfull graph, all nodes have probability $p$ of being connected. As previously explained, the simplest way to generate these edges involves flipping $n^2$ weighted coins or ``coin-flipping.'' The problem with coin-flipping is that it is extremely inefficient when generating an \ERfull model of a real-world network. A common characteristic of real-world networks is that most pairs of edges do not exist. Just consider how many friends you have on Facebook or followers on Twitter. It is likely to be an extremely small fraction of the billions of people present on these social networks. This makes most of the resulting adjacency matrices entries equal to zero, which results in something called a \emph{sparse matrix}.  Coin-flipping 
\aside{note:sparse}{We say a matrix is \emph{sparse} if the number of zero entries is so large that it makes sense just to store information on the locations of the entries that are nonzero. }
is an expensive procedure for this scenario because it requires a coin flip for every conceivable pair of nodes given by each entry of the matrix. Our goal is something that does work proportional to the number of \emph{edges} of the resulting network.

\subsection{Ball-dropping}
\label{sec:ball_dropping}
\begin{figure}
\centering
\includegraphics[width=4in]{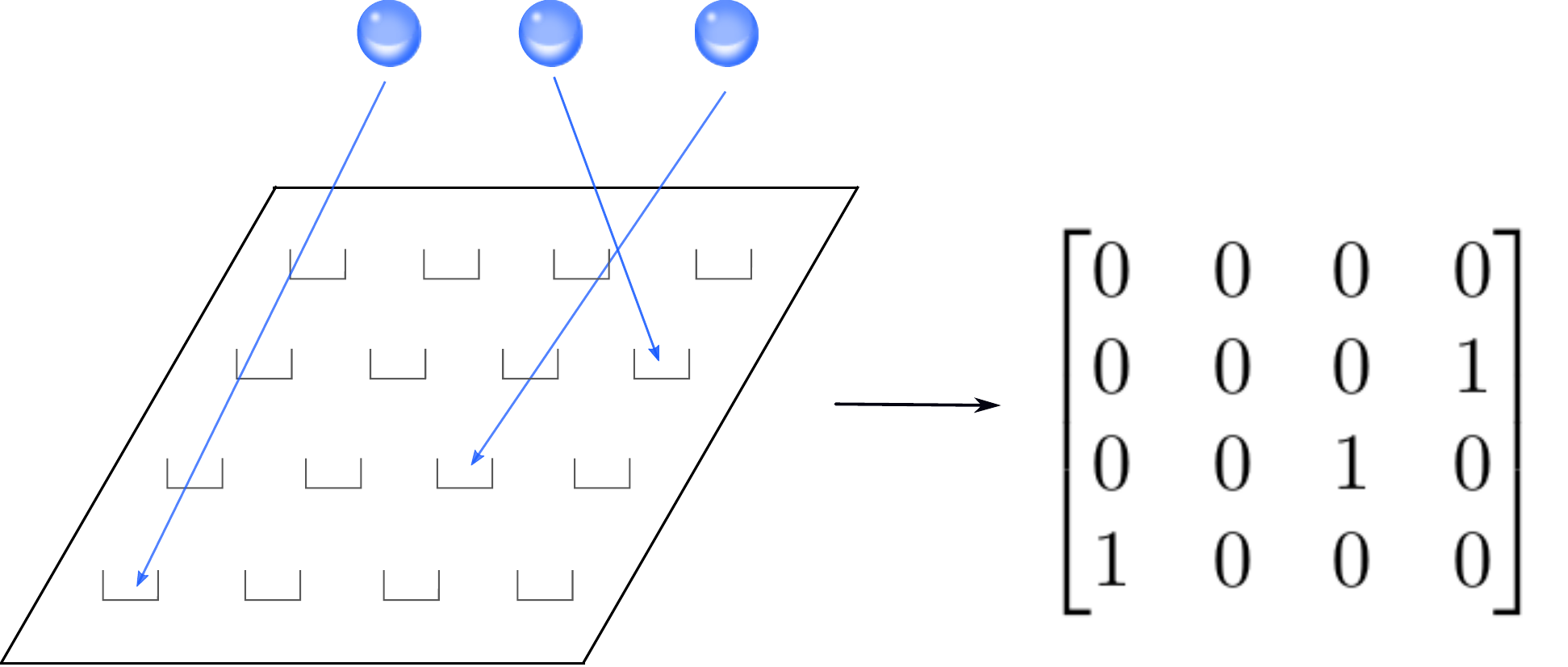}
\caption{A simple illustration of the ball-dropping procedure to generate an \ERfull graph. ``Balls'' are dropped with equal probability, and wherever balls are dropped, generate an edge in the adjacency matrix.}
\label{fig:balldrop}
\end{figure}

The first accelerated procedure we'll consider is what we call ball-dropping. The inspiration behind this method is that we can easily ``drop a ball'' into a uniform random entry in a large matrix, which results in a random edge, see figure \ref{fig:balldrop}. Careful and repeated use of this procedure will allow us to generate a random adjacency matrix where non-edges are never considered! This solves the efficiency problem that motivates our accelerated sampling procedures. 

The ball-dropping process, itself, is quite simple: 
\begin{compactenum}
\item generate a uniform random integer between $1$ and $n$ for $i$, 
\item generate a uniform random integer between $1$ and $n$ for $j$.
\end{compactenum}
The probability that we generate any entry $i,j$ is then $1/n \cdot 1/n = 1/n^2$, which is a uniform distribution over entries of the matrix. Two questions arise: (i) how many balls (or edges) do we drop and (ii) what do we do about duplicates? 

The solution to question (i) can be resolved by \emph{binomial random variables}. The number of balls that we want to drop is given by the number of edges we will generate in an \ERfull graph. Recall that each edge or entry in $A_{ij}$ is the result of a random, independent coin flip or Bernoulli trial. The total number of edges is the number of successes in these coin flips, which is exactly what a binomial random variable describes. Thus, the number of edges in an \ERfull graph is a binomial random variable with $n^2$ trials and probability $p$ of success. Many standard programming libraries include routines for sampling binomial random variables, which makes finding the number of balls to drop an easy calculation. 

\aside{aside:performance}{The approximate distribution that results from ignoring duplicate ball drops may still be useful if the random graph were used to test system performance or in other applications where the precise distribution is not required.} 
That leaves question (ii): what do we do about duplicate ball drops? Suppose the sample from the binomial distribution specifies $m$ edges. Each ball-drop is \emph{exactly} the procedure described above and we make no provisions to avoid duplicate entries. One strategy would be to ignore duplicate ball drops. While this is expedient, it samples from a different distribution over graphs than \ERfull, since less than $m$ unique edges will likely be generated.
The alternative is to discard duplicate ball drops and continue dropping balls until we have exactly $m$ distinct entries. This method gives the correct \ERfull distribution, and listing~\ref{code:ball-drop} implements this procedure. This procedure returns the edges of a random graph instead of the adjacency matrix to support the \emph{sparse} use case that motivates our accelerated study. 

\begin{listing}
\caption{A simple code to generate an \ERfull random graph by ball dropping}
\label{code:ball-drop}
\begin{minted}[fontsize=\footnotesize]{python}
import random      # random.randint(a,b) gives a uniform int from a to b
import numpy as np # numpy is the Python matrix package
"""
Generate a random Erdos-Renyi graph by ball-dropping. The input is:
   n: the number of nodes
   p: the probability of an edge
The result is a list of directed edges. 

Example:
  ball_drop_er(8,0.25) # 8 node Erdos-Renyi with probability 0.25
"""
def ball_drop_er(n,p):
    m = int(np.random.binomial(n*n,p))      # the number of edges
    edges = set()                           # store the set of edges
    while len(edges) < m:
        # the entire ball drop procedure is one line, we use python indices in 0,n-1 here
        e = (random.randint(0,n-1),random.randint(0,n-1))
        if e not in edges:                  # check for duplicates
            edges.add(e)                    # add it to the list
    return list(edges)                      # convert the set into a list
\end{minted}
\end{listing}

\paragraph{Proof that this procedure exactly matches the \ERfull description} We asserted above that this procedure actually generates an \ERfull graph where each edge $i,j$ occurs with probability $p$. This is easy to prove and is well-known, although proofs can be difficult to find. We follow the well-written example by~\citet{moreno}. In order for an edge $i,j$ to occur, it must occur after we have picked $m$, the number of edges from the binomial. Thus,
\begin{equation}
\text{Prob}[A_{ij} = 1] = \sum_{m=0}^{n^2} \text{Prob}[A_{ij} = 1 \mid \text{$\mA$ has $m$ edges}] \cdot \text{Prob}[\text{$\mA$ has $m$ edges}].
\end{equation}
The first term, $\text{Prob}[A_{ij} = 1 \mid \text{$\mA$ has $m$ edges}]$ is equal to $m/n^2$ because we are sampling without replacement. The second term $\text{Prob}[\text{$\mA$ has $m$ edges}]$ is exactly a binomial distribution. Hence, 
\begin{equation}
\begin{aligned}
\text{Prob}[A_{ij} = 1] & = \sum_{m=0}^{n^2} (m/n^2) \cdot \binom{n^2}{m} p^m (1-p)^{n^2 -m}\\
& = (1/n^2) \underbrace{\sum_{m=0}^{n^2} m \binom{n^2}{m} p^m (1-p)^{n^2 -m}}_{= \text{ expectation of binomial}} = (1/n^2) \cdot n^2 p = p.
\end{aligned}
\end{equation}
\par \vspace{-\baselineskip}
\QED

The downside with this method is that we do not have a precise runtime because the algorithm continues to generate edges until the number of unique edges is exactly $m$. We analyze this case further in Section~\ref{sec:er-analysis}.

\subsection{Grass-hopping}
\label{sec:grass-hopping}
\begin{figure}
\centering
\includegraphics[width=.4\textwidth]{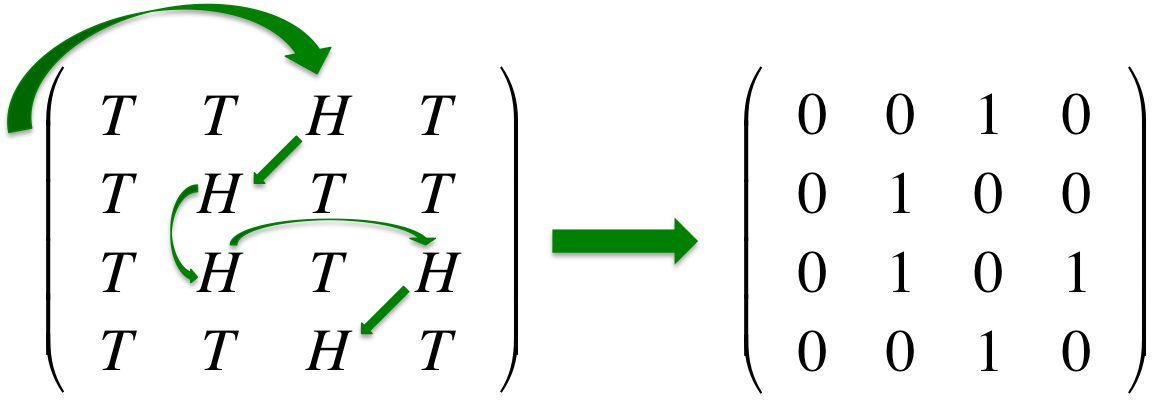}
\caption{An illustration of grass-hopping. The idea with grass-hopping is that we can sample geometric random variables to move between the coin-flips that produce edges (the heads) directly and skip over all the tails.} 

\label{fig:grass-hopping}
\end{figure}

The second accelerated method we present is what we have called \emph{grass-hopping}. This method is also known, but hard to find described (and we still encounter many individuals unaware of it!). The idea dates back to the 1960s (see Section~\ref{sec:fast-er-history}).
We first learned of it while studying the source code for the Boost Graph Library implementation of \ERfull graph generation~\cite{siek2001-bgl}. The essence of the idea is that we wish to ``grass-hop'' from edge to edge as we conceptually (but not actually) flip all $n^2$ coins in the coin flipping method. Note that in the real-world case, there will be many coin flips that come up as ``no edge'' repeated. Ideally, we'd like to ``hop'' over all of these flips as illustrated in Figure~\ref{fig:grass-hopping}. Is such a task possible?

Indeed it is. For a Bernoulli random variable with probability $p$, the number of consecutively failed trials can be derived from a geometric random variable. More specifically, a geometric random variable models the number of trials needed to obtain the next success in a series of coin flips with fixed probability. Thus, by sampling a geometric random variable, we can ``grass-hop'' from success-to-success and skip over all of the failed trials. A code implementing this is given in Listing~\ref{code:grass-hop}.

\paragraph{Some useful properties of geometric random variables}
A geometric random variable is a discrete random variable parameterized by the value $p$, which is the success probability of the Bernoulli trial. Let $X$ be a geometric random variable with probability $p$, the probability distribution function $\text{Prob}[X = k] = (1-p)^{k-1} p$. Note that this is just the probability that a coin comes up tails $k-1$ times in a row before coming up heads in the last trial. The expected value of $X$ is $1/p$. A straightforward calculation provides the variance as $\frac{1-p}{p^2}$.

\begin{listing}
\caption{A simple code to generate an \ERfull random graph by grass-hopping}
\label{code:grass-hop}
\begin{minted}[fontsize=\footnotesize]{python}
import numpy as np # numpy is the Python matrix package and np.random.geometric
                   # is a geometric random variate generator
"""
Generate a random Erdos-Renyi graph by grass-hopping. The input is:
   n: the number of nodes
   p: the probability of an edge
The result is a list of directed edges. 

Example:
  grass_hop_er(8,0.25) # 8 node Erdos-Renyi with probability 0.25
"""
def grass_hop_er(n,p):
    edgeindex = -1                   # we label edges from 0 to n^2-1
    gap = np.random.geometric(p)     # first distance to edge
    edges = []
    while edgeindex+gap < n*n:       # check to make sure we have a valid index
        edgeindex += gap             # increment the index
        src = edgeindex // n         # use integer division, gives src in [0, n-1]
        dst = edgeindex - n*src      # identify the column 
        edges.append((src,dst))
        gap = np.random.geometric(p) # generate the next gap
    return edges
\end{minted}
\end{listing}

\paragraph{Proof that grass-hopping is correct}
The proof that grass-hopping is correct is essentially just the result that the geometric random variable models the gaps between successes. However, in the interest of exposition, we present a proof that $\text{Prob}[A_{ij} = 1] = p$ explicitly. In the adjacency matrix of the graph, the probability that the $i,j$ cell will have an edge can be calculated by adding the probabilities of all sequences of hops that land on that cell. Let $\ell$ be the index of the $i,j$ cell in the linear order from $0$ to $n^2-1$ used in the code. The probability that $A_{ij} = 1$ is then the probability that the sequence of hops lands on $\ell$ or equivalently, is simply a series of geometric random variables whose sum is equal to $\ell+1$. (Here, we have an \emph{off-by-one} change because the indices range from $0$ to $n^2-1$ but the sums of gaps range from $1$ to $n^2$.) 

From the total probability theorem, the probability that an edge will be generated is the sum of the probabilities of all length-$k$ hop paths where $k$ goes from 1 to $\ell$+1.
\begin{equation}  \label{eq:path-expansion}
\begin{aligned}
\text{Prob}[A_{\ell}=1]&=\text{Prob}[A_{\ell}=1 \text{ in 1 hop}] + \text{Prob}[A_{\ell}=1 \text{ in 2 hops}] + \cdots \\
& \qquad + \text{Prob}[A_{\ell}=1 \text{ in $\ell +1$ hops}]\\
\end{aligned}
\end{equation}
There are a number of ways to land on index $l$ by $k$ hops. For example, here are two different length $k = 4$ hops landing on index $l = 13$:
\begin{center}
\scalebox{0.7}{
\begin{tikzpicture}[scale=0.28]
\renewcommand*{\VertexInterMinSize}{8pt}
\renewcommand*{\VertexSmallMinSize}{0pt}
\Vertex[x=0,y=3]{-1}
\Vertex[x=3,y=3]{0}
\Vertex[x=6,y=3]{1}
\Vertex[x=9,y=3]{2}
\Vertex[x=12,y=3]{3}
\Vertex[x=15,y=3]{4}
\Vertex[x=18,y=3]{5}
\Vertex[x=21,y=3]{6}
\Vertex[x=24,y=3]{7}
\Vertex[x=27,y=3]{8}
\Vertex[x=30,y=3]{9}
\Vertex[x=33,y=3]{10}
\Vertex[x=36,y=3]{11}
\Vertex[x=39,y=3]{12}
\Vertex[x=42,y=3]{13}

\SetUpEdge[style={post,bend left}]
\Edge(-1)(2)
\Edge(2)(5)
\Edge(5)(9)
\Edge(9)(13)
\end{tikzpicture}
}
\\
\scalebox{0.7}{\begin{tikzpicture}[scale=0.28]
    \renewcommand*{\VertexInterMinSize}{8pt}
  \renewcommand*{\VertexSmallMinSize}{0pt}
\Vertex[x=0,y=3]{-1}
\Vertex[x=3,y=3]{0}
\Vertex[x=6,y=3]{1}
\Vertex[x=9,y=3]{2}
\Vertex[x=12,y=3]{3}
\Vertex[x=15,y=3]{4}
\Vertex[x=18,y=3]{5}
\Vertex[x=21,y=3]{6}
\Vertex[x=24,y=3]{7}
\Vertex[x=27,y=3]{8}
\Vertex[x=30,y=3]{9}
\Vertex[x=33,y=3]{10}
\Vertex[x=36,y=3]{11}
\Vertex[x=39,y=3]{12}
\Vertex[x=42,y=3]{13}

\SetUpEdge[style={post,bend left}]
\Edge(-1)(0)
\Edge(0)(4)
\Edge(4)(6)
\Edge(6)(13)
\end{tikzpicture}
}
\end{center}
\noindent Let $q = 1-p$. Notice that the probability of the first path is 
\[ \underbrace{(q^2 p)}_{\text{hop 1}} \underbrace{(q^2 p)}_{\text{hop 2}} \underbrace{(q^3 p)}_{\text{hop 3}} \underbrace{(q^3 p)}_{\text{hop 4}} = q^{10} p^4. \]  The probability of the second path is $(p) (q^3 p) (q p) (q^6 p) = q^{10} p^4$. The fact that these two are equal is not a coincidence. In fact, the probability of a hop path with $k$ hops landing on $(\ell+1)$ is the same for any such path! There are $\binom{\ell}{k-1}$ such paths, and so the overall probability of a hop path with $k$ hops landing on $(\ell+1)$ is $\binom{\ell}{k-1}q^{\ell-k+1}p^{k-1}p$, where again $q=1-p$. We see this formally by observing that any hop path is a series of geometric random variables each with distribution $\text{Prob}[X=x]=q^{x-1}p$. Furthermore, all hop-paths with the same number of hops have the same probability of occurring because any sequence of $k$ geometric random variables will be an arrangement of $(\ell-k+1)$ $q$'s and $k$ $p$'s, where we will assert that the last letter is a $p$. Thus, the number of length $k$ hop-paths can be generated using binomial coefficients for the number of ways to arrange $(\ell-k+1)$ $q$'s and $k-1$ $p$'s.

On substituting these binomial coefficient probability expressions into \eqref{eq:path-expansion}:
\begin{equation}
\text{Prob}[A_{\ell}=1]= \underbrace{\textstyle \sum_{k=1}^{\ell+1}\binom{\ell}{k-1}q^{\ell-k+1}p^{k-1}}_{\text{binomial expansion of $(p+q)^\ell$}} {}\cdot p = p.
\end{equation}
Notice we are left with the binomial expansion of $(p+q)^{\ell} = 1$ multiplied by $p$. \QED

The runtime of this procedure is exactly $O(E)$ where $E$ is the number of edges output because for each instance of the loop, we generate one distinct edge. (This assumes a constant-time routine to generate a geometric random variable and an appropriate data-structure to capture their results.)

\subsection{Comparing ball-dropping to grass-hopping with coupon collectors}
\label{sec:er-analysis}
Both ball-dropping and grass-hopping produce the correct distribution, but our next question concerns their efficiency. For grass-hopping, this was easy. For ball-dropping, note that we need to drop at least $m$ balls to get $m$ edges. But we haven't yet discussed how many duplicate entries we expect. Intuitively, if $n$ is very large and $p$ is small, then we would expect few duplicates. Our goal is a sharper analysis.  Specifically, how many times do we have to drop balls in order get exactly $m$ distinct edges? 

This is a variation on a problem called the coupon collector problem. The classic coupon collector problem is that there are $m$ coupons that an individual needs to collect in order to win a contest. Each round, the individual receives a random coupon with equal probability. How many rounds do we expect before the individual wins? A closely related analysis to what we present shows that we expect the game to run for $m \log m$ rounds. 

In our variation of the coupon collector's problem, we must find the expected number of draws needed to obtain $m$ unique objects (these are edges) out of a set of $n^2$ objects (these are entries of the matrix) where we draw objects with replacement. This can be analyzed as follows based on a post from \url{math.stackexchange.com}~\cite{euyu-2012-coupon}.

Let $X$ be the random variable representing the number of draws needed to obtain the $m^{th}$ unique object. Let $X_i$ be a random variable representing the number of trials needed to obtain the $i^{th}$ unique object given that $i-1$ unique objects have already been drawn. Then, $E[X] = E[X_1] + E[X_2]+\ldots+E[X_m].$ Each random variable $X_i$ is geometrically distributed with $p = 1-\frac{i-1}{n^2}$ because the probability of drawing a new object is simply the complement of the chance of drawing an object that has already been drawn. Using the expected value of a geometric random variable, we obtain $E[X_i]=\frac{n^2}{n^2-i+1}$. Thus, $E[X]=\sum_{i=1}^{m} \smash{\frac{n^2}{n^2-i+1}}$. This can be rewritten in terms of harmonic sums. The harmonic sum, $H_n$, is defined as the sum of the reciprocals of the first n integers: $H_n=\sum_{i=1}^{n} \frac{1}{n}$. Consequently, $E[X] = n^2(H_{n^2} - H_{{n^2}-m})$. 
The harmonic partial sums can be a approximated as follows:  $H_n = \sum_{i=1}^{n} \frac{1}{i} \approx \log (n+1) + \gamma$~\cite{pollack}. \aside{note-em}{Here, the Euler-Mascheroni constant, $\gamma$, is approximately $0.57721$.
}
This approximation is derived from the left hand Riemann sum approximation for the integral of $\int_1^n \frac{1}{x}\, dx$.  The difference  between the Riemann sum, which represents the harmonic sum, and the integral, is the Euler-Mascheroni constant, $\gamma$. $\sum_{i=1}^{\infty} \frac{1}{i} - \int_{1}^{\infty} \frac{1}{x}=\gamma$.
After substituting in this approximation, the expected number of draws, $E[X]$, simplifies to $n^2 [ \log (n^2) - \log(n^2 - m) ]$. To get a better expression, we need a value of $m$. Recall that $m$ is sampled from a binomial whose expectation is $n^2 p$. Using $m = n^2 p$ simplifies the expression to: $E[X] = n^2 \log \tfrac{1}{1-p} = m p^{-1} \log \tfrac{1}{1-p}$. Thus, the extra work involved is $p^{-1} \log \tfrac{1}{1-p}$. In Problem~\ref{prob:ball-drop-oversampling}, we show that $p^{-1} \log \tfrac{1}{1-p} \ge 1$ with a removable singularity at $p=0$. 

\begin{figure}
\centering
\includegraphics[width=0.5\textwidth]{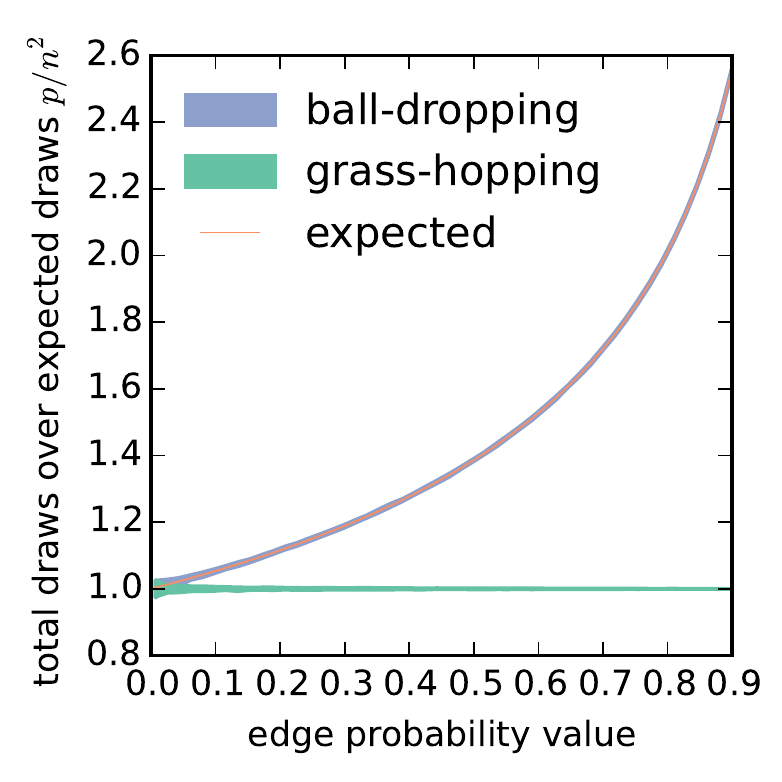}%
\includegraphics[width=0.5\textwidth]{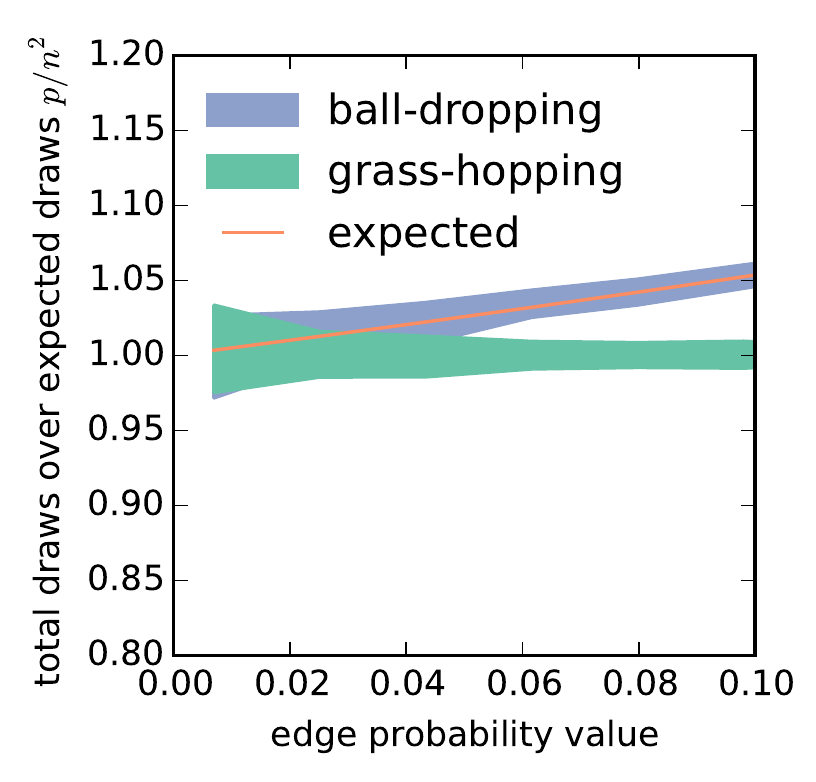}
\caption{At left, the number of random draws that the ball-dropping and grass-hopping  procedures must make as a function of $p$, normalized by the expected number $p n^2$. At right, the same data, but zoomed into the region where $p < 0.1$. For this figure, $n=1000$, and we show the 1\% and 99\% percentiles of the data in the region. The orange line plots the expected ratio $\tfrac{1}{p} \log( \frac{1}{1-p} )$.}
\label{fig:coupon}
\end{figure}

Consequently, ball-dropping always takes more random samples than grass-hopping. The difference between these two becomes considerable as $p$ gets larger as show in Figure~\ref{fig:coupon}. 

\paragraph{Large values of $p$} At this point, we should make the following observation. There is little reason to run ball-dropping once $p$ gets large. This is for two reasons. First, if $p$ is any value like $0.1$, then we would expect $0.1 n^2$ ``edges'', in which case we might as well do coin-flipping because it talks almost the same amount of time. So we generally expect $p$ to scale $O(1/n)$ or $1/n^\gamma$ for $\gamma < 1$, such as $\gamma = 1/2$.  In the models we will consider, however, there are often cases where $p$ does get large in small regions. Once $p > 0.5$, then we actually expect \emph{most} edges to be present in the graph. In which case it's more computationally efficient to use ball-dropping to determine which edges are \emph{missing} instead of which edges are present.

\subsection{Literature on efficient \ERfull generation: leap-frogging}
\label{sec:fast-er-history}

The idea of using a geometric distribution to grass-hop through the probabilities has been known since the 1960s~\cite{FanMullerRezucha1962}. In this original literature, it was called a ``leap-frog'' procedure.  But this is still not widely known and not always widely implemented -- despite this point being mentioned over a decade ago by~\citet{BatageljBrandes2005}, where it was called a geometric method. More recently, \citet{HagbergLemons2015} discuss the idea, but simply describe it as sampling from a waiting-time distribution. 




\section{Chung-Lu and Stochastic Block Models: Unions of \ERfull graphs}
\label{sec:chung-lu-union}
\label{sec:sbm-union}
Thus far, we have only studied fast methods for \ERfull graphs. The same techniques, however, apply to Chung-Lu and Stochastic Block Model graphs as well, because these probability matrices are \emph{unions of \ERfull blocks}. This is immediate for the stochastic block model because that model defines a set of small regions. For instance, it is easy to use our \ERfull subroutines to create a two block stochastic block model as the following code illustrates. 

\begin{minted}[fontsize=\footnotesize]{python}
""" Generate edges for a stochastic block model with two blocks.
n1 and n2 are the sizes of the two blocks and p is the within group
probability and q is the between group probability. 
Example: sbm(20,15,0.5,0.1) # 20 nodes in group 1, 15 nodes in group 2, ...
"""
def sbm2(n1,n2,p,q):
  edges = grass_hop_er(n1,p)                             # generate the n1-by-n1 block
  edges.extend( [ (i+n1,j+n1) for i,j in grass_hop_er(n2,p) ])        # n2-by-n2 block
  edges.extend( [ (i,j+n1) for i,j in grass_hop_er(max(n1,n2),q) if i < n1 and j < n2])
  edges.extend( [ (i+n1,j) for i,j in grass_hop_er(max(n1,n2),q) if i < n2 and j < n1])
  return edges
\end{minted}
This program has a runtime that scales with the number of edges in the graph. 

For Chung-Lu graphs, the situation is slightly more intricate. The example Chung-Lu probability matrix $\mP$ from Section~\ref{sec:chung-lu} has 8 nodes, 16 different probabilities, and 64 total entries in the probability matrix. It turns out that the number of distinct probabilities is given by the number of distinct degrees. That example had four distinct degrees: 4, 3, 2, 1. Since the probability of each matrix entry is $P_{ij} = d_i d_j / \text{total-degree}$ (where total-degree $ =  \sum_i d_i$ is just a constant), then if there are $t$ distinct degrees, we will have $t^2$ distinct probabilities in $\mP$. This corresponds with $t^2$ \ERfull blocks. This can be justified most easily if the vector of degrees $\vd$ is sorted. Let $d_{[1]}$ be the first degree in sorted order and $d_{[2]}$ be the next, and so on. Then the entire vector is: 
\begin{equation}
\vd = [ \, 
\underbrace{d_{[1]} \; d_{[1]} \; \cdots \; d_{[1]} }_{n_1 \text{ times}} \;  
\underbrace{d_{[2]} \; d_{[2]} \; \cdots \; d_{[2]} }_{n_2 \text{ times}} \;
\cdots \; 
\underbrace{d_{[t]} \; d_{[t]} \; \cdots \; d_{[t]} }_{n_t \text{ times}}
\, ].
\end{equation}
It does not matter if the vector is sorted in increasing or decreasing order and many of the values of $n_i$ may be $1$. Let $\rho = \text{total-degree}$. The resulting matrix $\mP$ now has a clear block \ERfull structure: 
\begin{equation}
\mP = 
\sbox0{$\begin{matrix}1&2&3\\0&1&2 \end{matrix}$}
  \begin{blockarray}{ccccc}
    &  n_1 & n_2 &  & n_t \\        
    \begin{block}{c[c|c|c|c]}       
n_1 & \vphantom{\usebox{0}}\makebox[\wd0]{$\frac{d_{[1]} d_{[1]}}{\rho}$} & \makebox[\wd0]{$\frac{d_{[1]} d_{[2]}}{\rho}$}  & \makebox[\wd0]{\large $\cdots$}& \makebox[\wd0]{$\frac{d_{[1]} d_{[t]}}{\rho}$} \\\cline{2-5}      
n_2 & \vphantom{\usebox{0}}\makebox[\wd0]{$\frac{d_{[2]} d_{[1]}}{\rho}$} & \makebox[\wd0]{$\frac{d_{[2]} d_{[2]}}{\rho}$} & & \\ \cline{2-5}    
    & \vphantom{\usebox{0}}\makebox[\wd0]{\large $\vdots$} & \makebox[\wd0]{\large $\vdots$} & \makebox[\wd0]{\large $\ddots$}&\\ \cline{2-5}
n_t & \vphantom{\usebox{0}}\makebox[\wd0]{$\frac{d_{[t]} d_{[1]}}{\rho}$} & \makebox[\wd0]{$\frac{d_{[t]} d_{[2]}}{\rho}$} & & \makebox[\wd0]{$\frac{d_{[t]} d_{[t]}}{\rho}$} \\        
  \end{block}
  \end{blockarray}.
\end{equation}
We leave an implementation of this as a \emph{solved} exercise at the conclusion of this paper. (Note that a very careful implementation can use $\binom{t}{2}$ \ERfull blocks.)  The first reference we are aware of to this algorithm for Chung-Lu graphs is~\citet{MillerHagberg2011}.

\paragraph{Direct Ball-dropping for Chung-Lu and Stochastic Block Models.} It turns out that there are equivalent procedures to ball dropping for both Chung-Lu and and Stochastic Block Models. We've deferred these to the problem section (Problems~\ref{prob:chung-lu-ball-drop},~\ref{prob:sbm-ball-drop}) where we explain some of the intuition behind the differences and some of the advantages and disadvantages. Both methods need to deal with the same type of duplicate entries that arise in ball-dropping for simple \ERfull models, however this setting is far more complicated as the different probability values can result in regions where there \emph{will} be many duplicates. These can be discarded to easily generate a slightly biased graph. This is what is done in the Graph500 benchmark, for example~\cite{Murphy-2010-graph500}. On the other hand, the number of \ERfull blocks grows quadratically for both models: $\binom{t}{2}$ for Chung-Lu and $k^2$ for the block model.  Thus, in the case where there are a large number of blocks, there are some non-trivial considerations as far as when grass-hopping would be preferable to ball-dropping. As an example, \citet{HagbergLemons2015} shows a technique that generalizes Chung-Lu and enables efficient sampling in this scenario.


\section{Fast Sampling of Kronecker Graphs via Grass-hopping}
\label{sec:fastkron}
As far as we are aware, a ball-dropping procedure or coin-flipping procedure was the standard method to sample a large Kronecker graph until very recently. Ball dropping methods often generated many duplicate edges in hard to control ways~\cite{Groer-2011-rmat}. In practice, these duplicates were often ignored to generate approximate distributions that are usable in many instances~\cite{Murphy-2010-graph500}. As previously mentioned, coin-flipping could never scale to large graphs. This situation changed when \citet{moreno} showed that a ``small'' number of \ERfull blocks were hiding inside the structure of the Kronecker matrix. 

We will walk through a new presentation of these results that makes a number of connections to various subproblems throughout discrete mathematics. The key challenge is identifying the \emph{entries of the matrix} where the \ERfull blocks occur. This was simple for both Chung-Lu and the stochastic block model, but it is the key challenge here. We are not going to discuss ball dropping procedures for Kronecker graphs because those are widely discussed elsewhere and it would be a diversion, but we strongly encourage interested readers to study both \citet{Chakrabarti-2004-rmat} and \citet{Seshadhri-2013-kronecker}. 

\subsection{Kronecker graphs as unions of \ERfull} \emph{And why grass-hopping is hard!}
You might be wondering what makes this problem hard once we exhibit that Kronecker graphs are unions of \ERfull regions. Looking at the matrix $\mP$ from Section~\ref{sec:kron} shows that there are some repeated probabilities. (Figure~\ref{fig:kron-regions-small} shows a guide to where the probabilities are the same.) However, it seems as if they are scattered and not at all square like they were in the Chung-Lu and Stochastic Block Model cases. There is a larger example in Figure~\ref{fig:one-big-kron-region} that shows what a \emph{single} region looks like for a $64$-by-$64$ matrix. This seems to imply that it is extremely difficult to take advantage of any \ERfull subregions. Moreover, it is unclear how many such regions there are. If there are many such regions that are all small, then it will not help us to sample each region quickly (even if we can!).

We are going to address both of these questions: \textit{(i)} How many \ERfull regions are there? and \textit{(ii)} How can each region be identified? More specifically, let the initiator matrix $\mK$ be $n$-by-$n$, and let $k$ be the number of Kronecker terms, so there are $n^k$ nodes in the graph. In the remainder of Section~\ref{sec:fastkron}, we show the following results:
\begin{compactitem}
\item \emph{Question (i).} There are $\binom{k+n^2-1}{k}$ regions, which is asymptotically $O(k^{n^2-1})$. This is important because it means that the number of regions grows more slowly than as the number of \emph{nodes} $n^k$, enabling us to directly enumerate the set of regions and keep an efficient procedure. (Section~\ref{sec:count_regions})
\item \emph{Question (ii).} The regions are easy to identify in a multiplication table view of the problem and the regions have a 1-1 correspondence with length $k$ non-decreasing sequences of elements up to $n^2$. (Section~\ref{sec:non-decreasing-sequences})
\item \emph{Question (ii).} We can randomly sample in the multiplication table view by unranking multiset permutations. (Section~\ref{sec:unrank-multiset})
\item \emph{Question (ii).} We can map between the multiplication table view and the repeated Kronecker products via Morton codes. (Section~\ref{sec:morton})
\end{compactitem}

\newsavebox{\smlmata}
\savebox{\smlmata}{$\sbmat{0.99 & 0.5 \\ 0.5 & 0.2}$}
\newsavebox{\smlmatc}
\savebox{\smlmatc}{$\sbmat{a & b \\ b & d}$}
\newsavebox{\smlmatb}
\savebox{\smlmatb}{$\sbmat{a & b \\ c & d}$}
\begin{figure}[tp]
\centering
\includegraphics[width=0.8\linewidth]{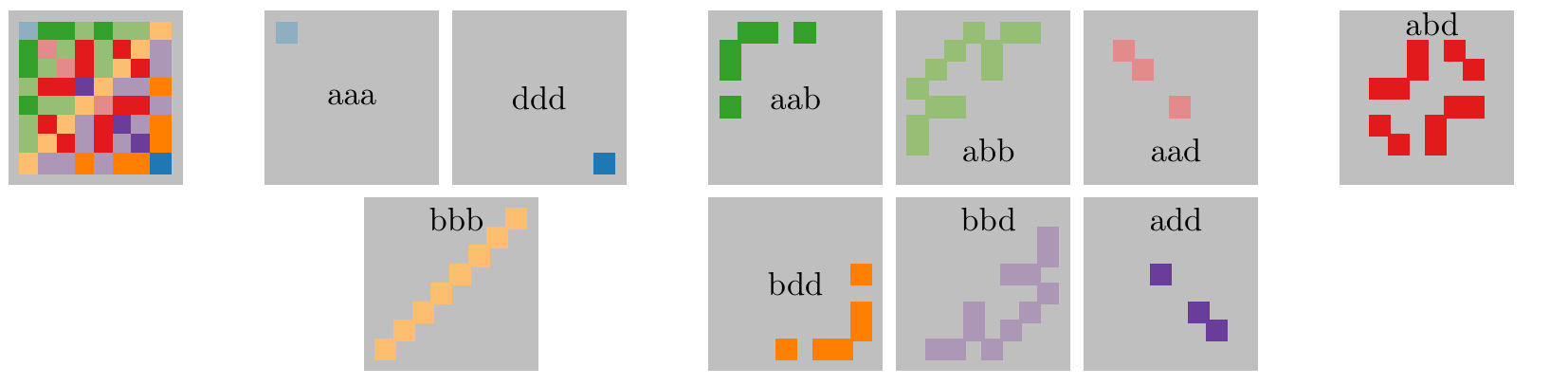}
\caption{For a Kronecker graph with a $2 \times 2$ initiator matrix $\mK = \usebox{\smlmata} = \usebox{\smlmatc}$ that has been ``$\kron$-powered'' three times to an $8 \times 8$ probability matrix, see equation~\eqref{eq:kron-example}, the marked cells illustrate the regions of identical probability that constitute an \ERfull piece. Here, we have used \emph{symmetry in $\mK$} to reduce the number of regions.}
\label{fig:kron-regions-small}
\savebox{\smlmata}{$\sbmat{0.99 & 0.5 \\ 0.5 & 0.2}$}\bigskip

\includegraphics[width=0.8\linewidth]{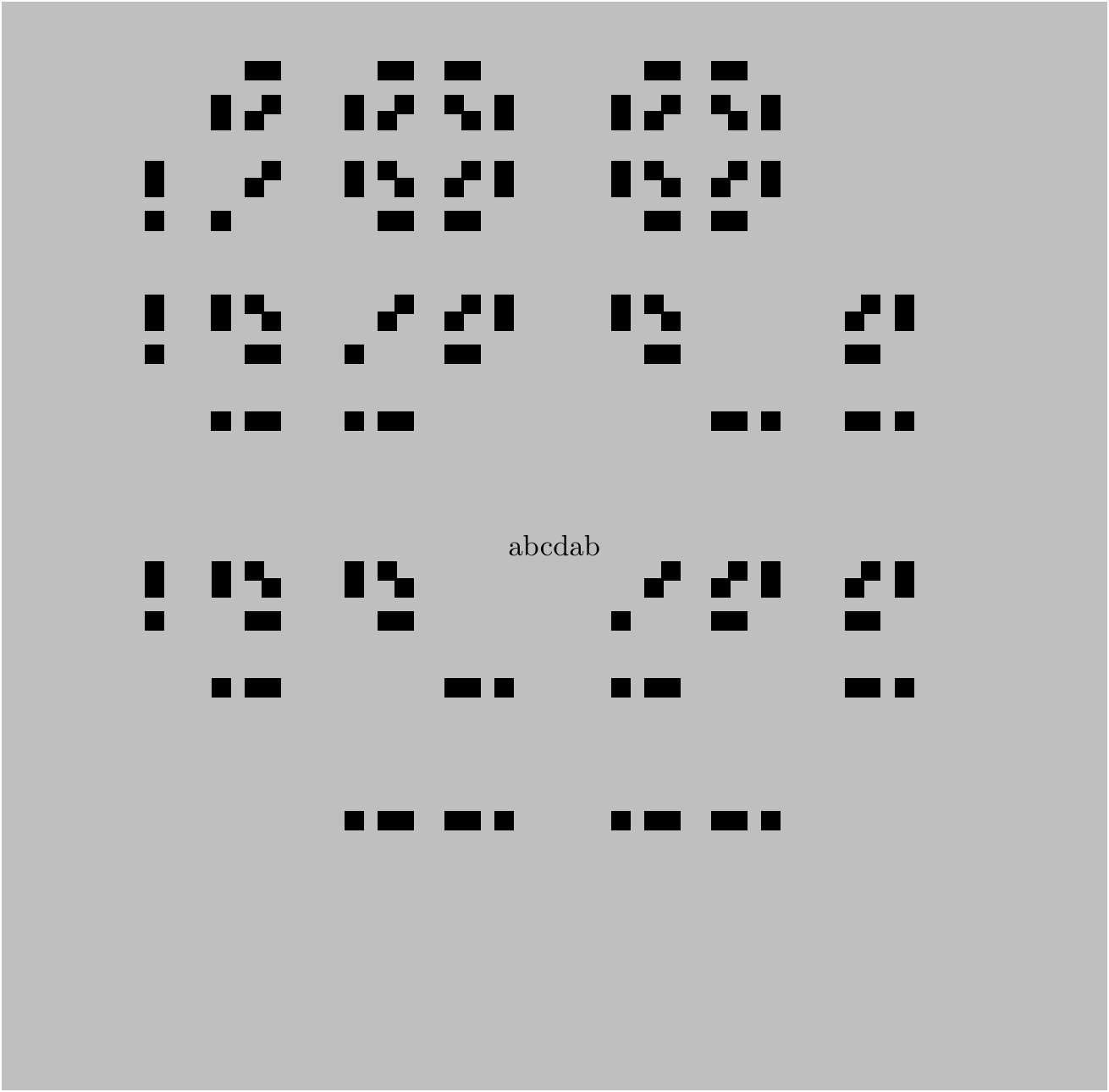} 
\caption{For a Kronecker graph with a $2 \times 2$ initiator matrix $\mK = \usebox{\smlmatb}$ that has been ``$\kron$-powered'' six times ($k=6$) to an $64 \times 64$ probability matrix, the marked cells illustrate one \ERfull piece that corresponds to the probability $abcdab$.}
\label{fig:one-big-kron-region}
\end{figure}


\subsection{The number of \ERfull regions in a Kronecker matrix}
\label{sec:count_regions}

Consider the result of taking a Kronecker product of a matrix $\mK$ with itself. In the $2 \times 2$ case with $\mK = \sbmat{a & b \\ c & d}$, then 
\begin{equation} \label{eq:kron-4}
\mK \kron \mK = 
\begin{bmatrix} aa&ab&ba&bb \\ ac&ad&bc&bd \\ ca&cb&da&db \\ cc&cd&dc&dd \end{bmatrix}.
\end{equation}
Note that all combinations of symbols $a,b,c,d$ occur between the Kronecker product. This result holds generally. Consequently, we can identify a length-$k$ string of the symbols $a,b,c,d$ with each entry of $\mP = \mKk$. If $\mK$ is $n \times n$, then there are $(n^2)^k$ such strings. The feature we are trying to exploit here is that the string $aba$ results in the same probability as $aab$ and $baa$, and hence we want to count the number of these distinct probabilities. 

An appropriate mathematical object here is the multiset. In a multiset, we have a set of items, along with a multiplicity, but where the order of items is irrelevant. So the strings $aab$ and $aba$ would both correspond to the multiset where $a$ occurs twice and $b$ occurs once. The cardinality of this multiset is $3$. Consequently, the number of unique probabilities in $\mKk$ is equivalent to the number of distinct multisets we can derive from the $n^2$ symbols in $\mK$ where each multiset has cardinality $k$. 

A famous and well-known result in counting follows. Recall that $\binom{n}{k}$ is the number of distinct sets of cardinality $k$ that can be drawn from a set of $n$ items.
The generalization to multisets is called ``multichoose'' and
\begin{equation}
\begin{aligned}
\left(\!\!\!\binom{n}{k}\!\!\!\right) & = \text{the number of multisets of cardinality $k$ with $n$ items} \\ & = \binom{n+k-1}{k}.
\end{aligned}
\end{equation}
The right hand side of this can be derived from a stars and bars argument (see, for instance, Wikipedia). 

Consequently, there are $\bigl(\mkern-5mu\binom{n^2}{k}\mkern-5mu\bigr) = \binom{k+n^2-1}{k}$ \ERfull regions in the probability matrix for a Kronecker graph. 

Recall that for large networks, we are expecting the network to be sparse, and so we expect the number of edges generated to be about the same size as the number of nodes. This expression, at the moment, is worrisome because there are $n^k$ nodes in a Kronecker graph and so we'd like there to be \emph{fewer} \ERfull regions than there are nodes. If this isn't the case, then just looking at all regions would result in more work than we would do generating edges within each region.

\aside{note:big-O}{We are using big-O notation here to understand what happens for large values of $k$ when $n$ is considered a fixed constant.}
We now show three results that will guarantee that there are fewer \ERfull regions than nodes of the graph for sufficiently large values of $k$. The first result is that there are at most $(e+1)^{k}$ regions for a large enough $k$, which gives our result when $e+1 \le n$. The second is that if $n = 2$ or $n=3$ there are at most $2^k$ or $3^k$ regions when $k \geq 10$. The third is that $\binom{n^2+k-1}{k}$ is $O(k^{n^2-1})$ asymptotically, which means that the number of regions asymptotically grows much more slowly than the number of nodes.  

\emph{Result 1.} By Sterling's approximation for combinations: $\binom{a}{b} \leq (\frac{e a}{b})^b$ \cite{Das-binom}. Substituting the expression for the number of regions yields: $\binom{k+n^2-1}{k} \leq (\frac{e(k+n^2-1)}{k})^k$. Once $k \ge e(n^2-1)$, then we have at most $(e+1)^k$ regions.

\emph{Result 2.} For $n=2$, we will show by induction that $\binom{k+n^2-1}{k} \leq 2^k$ when $k$ becomes large. Note $\binom{k+n^2-1}{k}=\binom{k+n^2-1}{n^2-1}$, so when $k=7$, $\binom{k+3}{3}=120$, $2^k=128$, and $120<128$. For our inductive step we must show that our inequality holds true for $k+1$ assuming it holds true for $k$. Substituting $k+1$, we need to show that  $\binom{k+4}{3} \leq 2^{k+1}$. 
Expanding the left hand side,
\begin{eqnarray}
\binom{k+4}{3}&=&\frac{(k+4)(k+3)(k+2)}{3!} \nonumber \\
&=& \frac{k+1}{k+1} \cdot \frac{(k+4)(k+3)(k+2)}{3!} \nonumber \\
&=& \frac{k+4}{k+1} \binom{k+1}{3} \nonumber \\
&\leq& 2 \cdot 2^k = 2^{k+1} \nonumber 
\end{eqnarray}

The last statement assumes $k \geq 2$, so $\frac{k+4}{k+1} \leq 2$.

The same strategy works for $n=3$ to show that $\binom{k+n^2-1}{k} \leq 3^k$ for $n=3$ when $k$ becomes large. When $k=10$, $\binom{k+8}{k}=43758$, $3^k=59049$, and $43758<59349$. For our inductive step we must show that our inequality holds true for $k+1$ assuming it holds true for $k$. Substituting $k+1$ into our inequality, we need to show  $\binom{k+9}{8} \leq 3^{k+1}$, or equivalently $\frac{\prod_{i=2}^{9} k+i}{8!} \leq 3^k \cdot 3$. Multiplying the LHS by $\frac{k+1}{k+1}$ allows us to use our inductive assumption that $\binom{k+8}{8} \leq 3^k$. Thus, we obtain $\frac{k+9}{k+1} \leq 3$ which is true when $k \geq 3$.

\emph{Result 3.} Finally, we note that $\binom{k+n^2-1}{k} = \frac{(k+n^2-1)!}{k! (n^2-1)!} = O(\frac{(k+n^2-1)!}{k!})$ because $n$ is a constant. If we use the simple upper-bound $\frac{(k+n^2-1)!}{k!} \le (k+n^2-1)^{n^2-1}$ and then take logs, we have: $(n^2 - 1) \log(k + n^2-1)$. The concavity of $\log$ gives the subadditive property $\log(a+b) \le \log(a)+\log(b)$ when $a, b \ge 2$. So $(n^2 - 1) \log(k + n^2-1) \le (n^2 -1) \log(k) + (n^2 - 1) \log(n^2-1)$. Exponentiating now yields $O(k^{n^2-1})$.

We now have the results that show there are sufficiently few \ERfull regions in a Kronecker graph, and so our grass-hopping procedure could successfully be applied to each region. This resolves question (i). We now turn to question (ii). 

\subsection{The strategy for grass-hopping on Kronecker graphs: Multiplication tables and Kronecker products}
In order to find the \ERfull regions in a Kronecker graph easily, we need to introduce another structure: the multiplication table. Let $\vv$ be the vector $\bmat{a & b & c & d}^T$, then 
\begin{equation}
\vv \kron \vv = \bmat{ a \vv \\ b \vv \\ c \vv \\ d \vv }.
\end{equation}
This object can be \emph{reshaped} into 
\begin{equation} \label{eq:mult-4}
\text{reshape}(\vv \kron \vv) = 
\bmat{ aa & ab & ac & ad \\
       ba & bb & bc & bd \\
       ca & cb & cc & cd \\ 
       da & db & dc & dd}
\end{equation}
For those accustomed to the mixed-product property of Kronecker products, this result is just $\tvec(\vv \vv^T) = \vv \kron \vv$. The \emph{vec} operator is a useful tool that converts a matrix into a vector by stitching the column vectors of a matrix together:
\begin{equation}
\begin{aligned}
\label{eqn:vec}
\text{vec}\left(\bmat{ a & b \\ c & d \\}\right) = \bmat{ a \\ c \\ b \\ d}
\end{aligned}
\end{equation}
The reason this is called a multiplication table is that it is exactly the multiplication table you likely learned in elementary school when $a = 1, b = 2, c = 3, d = 4, \ldots$. 

To get some sense of where this is going, compare~\eqref{eq:mult-4} to~\eqref{eq:kron-4} and notice that all the same entries occur, but they are just reorganized. 
The idea with a multiplication table is that we can keep going: 
\begin{equation}
\text{reshape}(\vv \kron \vv \kron \vv) = \vcenter{\hbox{\includegraphics[scale=0.8]{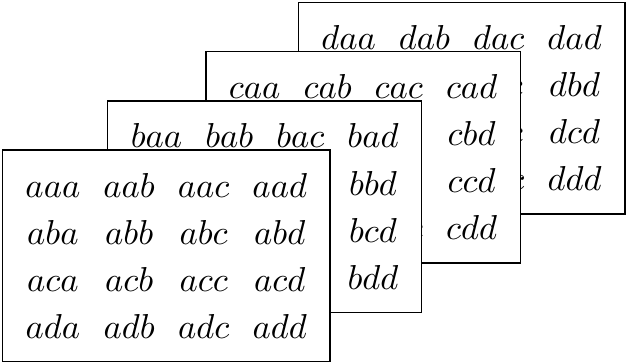}}}.
\end{equation}
More generally, a $k$-dimensional multiplication table arises from 
\begin{equation} 
\text{reshape}(\underbrace{\vv \kron \vv \kron \cdots \kron \vv}_{k \text{ times}}).
\end{equation} 

Here's how the multiplication table view helps us: \emph{it is easy to find the \ERfull regions in a multiplication table!} Consider the entry $aab$. This occurs in cells $(1,1,2)$, $(1,2,1)$, and $(2,1,1)$. These are exactly the three permutations of the multiset with two $1$s and one $2$. 

However, that result does not do us any good without the following fact: \emph{the entries of the $k$-dimensional multiplication table with $\vv = \text{vec}(\mK)$ can be mapped 1-1 to entries of the repeated Kronecker product matrix $\smash{\underbrace{\mK \kron \mK \kron \cdots \kron \mK}_{k \text{ times}}}$}. More specifically, we show that
\begin{equation}
\label{eq:morton-kron}
\tvec(\underbrace{\mK \kron \mK \kron \cdots \kron \mK}_{k \text{ times}}) = \mM_{n,k}\underbrace{\tvec(\mK) \kron \tvec(\mK) \kron \cdots \kron \tvec(\mK)}_{k \text{ times}}
\end{equation}
where $\mM_{n,k}$ is a permutation matrix based on a Morton code.  This mapping is illustrated in Figure~\ref{fig:morton-small} and Figure~\ref{fig:morton-big}. We will explain exactly what is in that figure in subsequent sections, but we hope it helps provide a visual reference for the idea of mapping between the Kronecker information and the multiplication table. 

These two results, together with the small number of regions result from Section~\ref{sec:count_regions}, enable the following strategy to grass-hop on an Kronecker graph efficiently. Grass-hop independently in each region of the the multiplication table and then map the multiplication table entries back to the Kronecker matrix.  More programmatically:
\begin{center}
\begin{minipage}{4in}
\begin{minted}[fontsize=\footnotesize]{python}
""" Generate a Kronecker graph via grass-hopping. The input K is the
Kronecker initiator matrix and the the value k is the number of levels. 
Example: grass_hop_kron([[0.99,0.5],[0.5,0.2]], 3) """
def grass_hop_kron(K,k):
  n = len(K) # get the number of rows 
  v = [K[i][j] for j in xrange(n) for i in xrange(n)] # vectorize by cols
  edges_mult = []
  for r in regions(v,k):            # for each region of the mult. table
    edges_mult.extend(grass_hop_region(r, v)) # get edges in mult. table
  edges_kron = []
  for e in edges_mult:              # map edges from mult. table to kron
    edges_kron.append(map_mult_to_kron(e, n))
  return edges_kron
\end{minted} 
\end{minipage}
\end{center}

The goal of the next three subsections is to write the each of the functions: \verb#regions# (Section~\ref{sec:non-decreasing-sequences}), \verb#grass_hop_region# (Section~\ref{sec:unrank-multiset}), and \verb#map_mult_to_kron# (Section~\ref{sec:morton}). Once we have these pieces, we 
will be able to efficiently grass-hop through a Kronecker graph! 

\subsection{Enumerating all \ERfull regions}
\label{sec:update}
\label{sec:non-decreasing-sequences}

Recall that each of the distinct probabilities that occur in $\mP = \mKk$ can be identified with a multiset of size $k$ from a collection of $n^2$ objects (Section~\ref{sec:count_regions}). Our task is to enumerate all of these multisets. Here, we note that each multiset can be identified with a non-decreasing sequence of length $k$ where the symbols are $0, 1, \ldots, n^2-1$. For instance, when $n=2, k=3$, these sequences are
\begin{multicols}{10}
\footnotesize
\noindent[0,0,0],\\~
 [0,0,1],\\~
 [0,0,2],\\~
 [0,0,3],\\~
 [0,1,1],\\~
 [0,1,2],\\~
 [0,1,3],\\~
 [0,2,2],\\~
 [0,2,3],\\~
 [0,3,3],\\~
 [1,1,1],\\~
 [1,1,2],\\~
 [1,1,3],\\~
 [1,2,2],\\~
 [1,2,3],\\~
 [1,3,3],\\~
 [2,2,2],\\~
 [2,2,3],\\~
 [2,3,3],\\~
 [3,3,3].
\end{multicols}
\noindent Let $\vv = \tvec(\sbmat{a & b \\ c & d}) = \sbmat{a & c & b & d}^T$.
The sequence $[0,2,2]$, then, refers to the probability $v(1)v(3)v(3) = a bb$. 

At this point, we are starting to mix programmatic indexes ($0, \ldots, n^2-1$) with the mathematical indices $(1, \ldots, n^2)$. Our goal is not to be confusing, but rather to ensure that the discussion in the text matches the programs more precisely from this point forward. 

As justification, we hope it is easy to see that each non-decreasing sequence of length $k$ corresponds with a multiset of cardinality $k$. (For instance, $[0, 2, 2, 3]$ corresponds to one $0$, two $2$s, and one $3$.) The other direction, that each multiset can be represented as a non-decreasing sequence, is also straightforward. Given a multiset of elements from $0$ to $n^2-1$ with $k$ total elements, place them into a sequence with repetitions in sorted order. So if we had two $5$s, three $1$s, and one $2$, we would get the sequence $[1, 1, 1, 2, 5, 5, 5]$. This sequence is non-decreasing and so we can do this for any multiset. 

\aside{note:duplicate}{One note, there could be fewer \ERfull regions due to duplicated probability values. We are assuming here that the values in $\mK$ are distinct.}
\emph{Thus, we have justified that the length $k$ non-decreasing sequences are in $1-1$ correspondence with the \ERfull regions.} We now continue with the problem of computationally enumerating non-decreasing sequences. 

For remainder of this section, $k$ will represent the length of the sequence, and $m$ will represent $\text{the largest entry} + 1$ that is valid in the sequence in order to match the code precisely. In the previous example, $k = 4$ and $m = 4$. Our code for this task is implemented in the subroutine \verb#regions# in Listing~\ref{code:update}. It begins with the first sequence: $[0, 0, 0, 0]$ and iteratively steps through all of them via the function \verb#next_region#. 

The function \verb#next_region# handles scenarios exemplified by the following three cases
\begin{equation}
\begin{aligned} 
[0, 1, 1, 2] & \to [0, 1, 1, 3] & & \text{easy} \\
[1, 3, 3, 3] & \to [2, 2, 2, 2] & & \text{spill} \\
[3, 3, 3, 3] & \to [-1,-1,-1,-1] & & \text{spill and done}.
\end{aligned}
\end{equation}
Recall that Python indexes from $0$. This \verb#next_region# update works by incrementing the last, or $k-1$th, entry of the sequence to the next value. In the easy case, this results in $[0, 1, 1, 2]  \to [0, 1, 1, 3]$. In the spill case, we get $[1, 3, 3, 3] \to [1, 3, 3, 4]$. We check for these cases by examining the value of this last element again. If it is any value $< m$, then we are in the easy case and there is nothing left to do. However, if the last value is equal to $m$, then we need to handle the spill. In the spill scenario, we recursively ask for next non-decreasing sequence for all but the last element. In our example, this call produces \verb#next_update([1, 3, 3], 4)#, which yields $[1, 3, 3] \to [2,2,2]$. At this point, \verb#cur = [2, 2, 2, 4]#. Because the sequence must be non-decreasing, we set the last element (currently at $4)$ to be the first value possible. This is given by the second-to-the-last element (2) after the prefix is updated the value is incremented by $1$. And so, this update produces $[1, 3, 3, 3] \to [2, 2, 2, 2]$. The final case is if the array had length $1$, then there is no prefix to update. In this case, we simply flag this scenario by introducing a $-1$ sentinel value. This sentinel value propagates through the array.

\begin{listing}
\caption{Code to update a non-decreasing sequence representing a subregion to the next non-decreasing sequence}
\label{code:update}
\begin{minted}[fontsize=\footnotesize]{python}
def next_region(cur, m):
  k = len(cur)
  cur[k-1] += 1      # increment the last element 
  if cur[k-1] == m:  # if there is spill 
    if len(cur) > 1: # there is an array
      cur[0:k-1] = next_region(cur[0:-1],m) # recur on prefix
      cur[k-1] = cur[k-2]                    # update suffix
    else:
      cur[k-1] = -1   # singleton, no room left! 
  return cur

""" Generate the set of Erdos-Reyni regions in a Kronecker graph
where v = vec(K), and k = number of levels. Each region gives indices
into v that produce a distinct Erdos-Reyni probability. 
Example: regions([0.99,0.5,0.5,0.2],3) """
def regions(v,k):
  m = len(v)
  rval = []
  cur = [ 0 for _ in xrange(k) ] # initialize the regions to zero
  while cur[0] != -1:
    rval.append(list(cur)) # make a copy
    next_region(cur, m)
  return rval
\end{minted}
\end{listing}

\emph{Proof that regions and update work.} The following is a sketch of a proof that \verb#next_region# gives the next region in lexicographic order. First, for arrays of length $1$, this is true because at each step we increment the element up until we generate the $-1$ termination symbol. We now inductively assume it is true for arrays of length $< k$. When \verb#next_region# runs on an array of length $k$, then either we increment the last element, in which case we have proven the result or the last element is already $n-1$. If the last element is already $n-1$, then we generate the next element in lexicographic order in the prefix array, which is of length $k-1$, and this occurs via our induction hypothesis. Finally, note that the last element of the updated prefix has the same value as our suffix.

\subsection{Grass-hopping within an \ERfull region in the Multiplication Table: Unranking multiset permutations}
\label{sec:unrank-multiset}

Given an \ERfull subregion, we now turn to the problem of how to grass-hop within that region in the multiplication table. Let $\vv$ be a length $m$ vector, which is $\tvec(\mK)$, and consider the $k$-dimensional multiplication table $\mMt$. The entry $(i,j, \ldots, l)$ in the multiplication table is simply equal to the product $v(i)v(j) \ldots v(l)$. In other words
\begin{equation}
\mMt(i,j,\ldots,\ell) = \underbrace{v(i) v(j) \cdots v(\ell)}_{\text{$k$ total terms}},
\end{equation}
where $v(i)$ is the $i$th entry in the vector $\vv$. 
Recall from section \ref{sec:update} that each \ERfull subregion exactly corresponds with a length $k$ non-decreasing sequence. Let $r$ be the length $k$ non-decreasing sequence that labels the current region. In the multiplication table, this region corresponds to the element 
\begin{equation} \label{eq:multi-element}
\mMt(r_1+1, r_2+1, \ldots, r_k+1) = v(r_1+1)v(r_2+1) \cdots v(r_k+1). 
\end{equation}
(Note that we index the region programmatically, but the vector entries mathematically, hence, the addition of 1.)
The locations in the multiplication table in which this \ERfull subregion occurs are the distinct permutations of the sequence $r$. For example
\begin{equation} \label{eqn:rperms}
\begin{aligned} 
r = [0,1,1,1] & \to && [0,1,1,1], [1,0,1,1], [1,1,0,1], [1,1,1,0] \\
r = [0,1,2,2] & \to && [0,1,2,2], [0,2,1,2],
 [0,2,2,1], [1,0,2,2], \\
 &&& [1,2,0,2], [1,2,2,0], [2,0,1,2], [2,0,2,1], \\ 
 &&& [2,1,0,2], [2,1,2,0], [2,2,0,1], [2,2,1,0].
\end{aligned}
\end{equation}
All of these entries have exactly the same probability because the multiplication operation in~\eqref{eq:multi-element} is associative. 

Consequently, it is easy to find the regions of the multiplication table that share the same probability. What is not entirely clear at this point is how we use this with a grass-hopping procedure. In grass-hopping, we move around indices of the \ERfull region in hops. How can we \emph{hop} from the element $[0,2,1,2]$ to $[2,0,1,2]$? 

The answer arises from an the \emph{ranking} and \emph{unranking} perspective on combinatorial enumeration~\cite{bonet}. For a complex combinatorial structure, such as multiset permutations, we can assign each permutation a rank from $0$ to the total number of objects minus one. Thus, in second example in (\ref{eqn:rperms}) we have 
\begin{equation} \label{eqn:rank_assignment}
\begin{aligned} 
[0,1,2,2] & \to 0 \\
[0,2,1,2] &\to 1 \\
& \vdots\\
 [2,2,1,0] & \to 11
\end{aligned}
\end{equation} 
This rank is based on the lexicographic order of the objects. Lexicographic order is exactly the same order you'd expect things to be in based on your experience with traditional sequences of numbers. The permutation $[2,1,0,2]$ occurs before $[2,2,0,1]$ for the same reason we say $2102$ is before $2201$ in order. \emph{Unranking} is the opposite process. Given an integer between $0$ and the total number of objects minus one, the \emph{unranking} problem is to turn that integer into the combinatorial object. Hence, 
\begin{equation}
4 \overset{\text{unrank}}{\to} [1,2,0,2].
\end{equation}
More generally, given the initial sequence $r = [0,1,2,2]$, and any integer between $0$ and $11$, the unranking problem is to turn that integer into the sequence with that rank in lexicographic order.

This is exactly what we need in order to do grass-hopping on multiset permutations.

We grass-hop on the integers between $0$ and the total number of multiset permutations $- 1$.  The total number of multiset permutations is given by the the formula:
\begin{equation}\label{eqn:num_perms}
\frac{m!}{a_1!  a_2! \ldots a_{k}!}
\end{equation}
where $a_i$ is the number of times the $i$th element appears and $m$ is the cardinality of the multiset. (See the permutation page on Wikipedia for more details)
At a high level, the unrank algorithm (Listing~\ref{code:unrank}) takes as input a multiset permutation in non-decreasing form represented as an array, as well as an index or ``rank'' of the desired permutation. The algorithm is recursively defined and simply returns the original array if the input rank is 0 as the base case. In the other case, the algorithm searches for the first element of the multiset permutation. (The details of this search are below.) Once it finds the first element, then it recurs on the remainder of the sequence after removing that element.

To make this code somewhat efficient, we maintain a counter representation of the multiset. This counts the number of times each element occurs in the multiset and keeps a sorted set of elements. For the multiset $[0, 1, 2, 2]$, the counter representation is: 
\begin{equation} 
\underbrace{\text{keys} = [0, 1, 2]}_{\text{sorted elements}} \text{ and } \textit{mset} = \begin{cases} 0 & \to 1 \\ 1 & \to 1 \\ 2 & \to 2 \end{cases} \end{equation}
The functions \verb#ndseq_to_counter# and \verb#counter_to_ndseq# convert between the counter and sequence representations. The function \verb#num_multiset_permutations# returns the number of permutations of a given multiset via~\eqref{eqn:num_perms}.
For our example, $[0, 1, 2, 2]$, equation (\ref{eqn:num_perms}) is equal to 12, which matches our direct enumeration in (\ref{eqn:rank_assignment}). 
The main algorithm is \verb#unrank_mset_counter#, which takes as input the counter representation of the multiset as well as the ``rank'' index. The objective of \verb#unrank_mset_counter# is to find the first element of the permutation in lexicographic order. We examine the sorted elements in the \verb#keys# array in order. For each element, we tentatively remove it by decreasing its count in \verb#mset# and then we count the number of multiset permutations with the remaining elements. For our running example, this yields: 
\begin{equation}
[0,1,2,2] \to 
\begin{array}{l} 
\text{starts with 0} \text{ and ends with a permutation of } [1,2,2] = 3 \\
\text{starts with 1} \text{ and ends with a permutation of } [0,2,2] = 3 \\
\text{starts with 2} \text{ and ends with a permutation of } [0,1,2] = 6. \\
\end{array}
\end{equation}
We interpret this output as follows. Let the desired rank be $R$. There are three permutations which start with a zero, and they correspond to the first three ranks 0, 1, or 2. So if $R < 3$, in that case we can recurse and unrank the permutation with sequence $[1,2,2]$ and exactly the same rank $R$. If the rank is 3, 4, or 5, then the permutation starts with $1$ and we can recurse and unrank the permutation of $[0,2,2]$ with rank $R - 3$. Finally, if the rank is $6, \ldots, 11$, then the permutation starts with $2$ and we can recurse and unrank the permutation of $[0,1,2]$ with rank $R-6$. These recursions yield the suffix that we return. 


\begin{listing}
\caption{Unranking a multiset permutation}
\label{code:unrank}
\begin{multicols}{2}
\begin{minted}[fontsize=\footnotesize]{python}
# unrank takes as input:
# - C: a multiset represented by a list
# - n: the lexicographic rank to find
# and returns the nth permutation of C in 
# lexicographic order.
#
# Examples:
# unrank([0,1,1,3], 0) returns [0,1,1,3]
# unrank([0,1,1,3], 1) returns [0,1,3,1]
# unrank([0,1,1,3], 2) returns [0,3,1,1]

from math import factorial 

def ndseq_to_counter(seq):
  mset = {}
  for c in seq:
    # get a value with a default
    # of zero if it isn't there
    mset[c] = mset.get(c,0)+1
  return mset, sorted(mset.keys())
  
def counter_to_ndseq(mset,keys):
  seq = []
  for k in keys: # keys in sorted order
    # append k mset[k] times
    for v in xrange(mset[k]):
      seq.append(k) 
  return seq

def num_multiset_permutations(mset):
  count = factorial(sum(mset.values()))
  for k in mset.keys():
    count = count//factorial(mset[k])
  return count
  
def unrank_mset_counter(mset,keys,n):
  if n==0:       # easy if rank == 0
    return counter_to_ndseq(mset,keys)
  for s in keys: # else find prefix key
    mset[s] -= 1 # decrease count of s
    # determine number of prefixes with s
    place = num_multiset_permutations(mset)
    if place > n:
      # then the first element is s
      if mset[s] == 0: # remove the key
        keys.remove(s) # if the count is 0
      suffix = unrank_mset_counter(
                mset, keys, n) # recurse!
      suffix.insert(0, s)      # append s
      return suffix
    else:  # here it does not start with s
      mset[s] += 1 # restore the count
      n -= place   # update search offset
  raise(ValueError("rank too large"))
  
def unrank(seq,n): 
  mset,keys = ndseq_to_counter(seq)
  return unrank_mset_counter(mset,keys,n)
\end{minted}
\end{multicols}
\end{listing}

\begin{listing}
\label{code:grass_hop}
\caption{Grass-hopping in an \ERfull region of a Kronecker product matrix}
\begin{minted}[fontsize=\footnotesize]{python}
import numpy as np # use np.random.geometric for the geometric random variables 
# grass_hop_region takes as input
# - r: the region to be sampled represented by a non-decreasing array      
# - v: the initiator matrix represented as a n^2-by-1 column vector
# and returns a list of edges represented by indexes in mult. table
# Example: v = [0.99,0.5,0.5,0.2]; grass_hop_region(regions(v,3)[2], v)
def grass_hop_region(r,v):
    p = multtable(r,v)               # p is the common prob value of the region
    n = num_multiset_permutations(ndseq_to_counter(r)[0]) # total size of region
    edges_mult = []                  # the initially empty list of edges 
    i = -1                           # starting index of the grass-hopping 
    gap = np.random.geometric(p)     # the first hop
    while i+gap < n:                 # check to make sure we haven't hopped out
        i += gap                     # increment the current index
        edges_mult.append(unrank(r,i))  # add the 
        gap = np.random.geometric(p) # generate the next gap
    return edges_mult

# multtable takes as input:
# - r: an array with k elements specifying a cell in the multiplication table
# - v: the initiator matrix represented as a n^2-by-1 column vector
# and returns the value at the specified location in the multiplication table
def multtable(r,v):
    final = 1.0
    for val in r:
        final *= v[val]
    return final
\end{minted}
\end{listing}
\subsection{Mapping from the Multiplication Table to the Kronecker graph: Morton codes}
\label{sec:morton}

In this section we prove a novel connection between repeated Kronecker product graph matrices and Morton codes that was mentioned in equation~\ref{eq:morton-kron}. But first, some background on Morton codes! 

Morton codes or Z-order codes are a neat primitive that arise in a surprising diversity of applications. Relevant details of these applications are beyond the scope of our tutorial, but they include high-performance computing \cite{Buluc-2009-sparse-morton}, database search, and computer graphics (see Wikipedia for references on these). When Morton codes and Z-order are used in the context of matrices, the term refers to a specific way to order the matrix elements when they are represented as a data array with a single index. Common strategies to accomplish this include \emph{row} and \emph{column} major order, that order elements by rows or columns, respectively.  Morton codes adopt a recursive, hierarchical pattern of indices. Here are four different ways to organize the $16$ values in a $4$-by-$4$ matrix 
\begin{equation*}
\underbrace{%
\mbox{%
\begin{tikzpicture}[ampersand replacement=\&]
\matrix (A) [matrix of math nodes,inner sep=0pt,left delimiter={[},right delimiter={]},column sep={8pt}, row sep={8pt}] 
{ 0 \& 1 \& 2 \& 3 \\
  4 \& 5 \& 6 \& 7 \\
  8 \& 9 \& 10 \& 11 \\
  12 \& 13 \& 14 \& 15 \\ 
};
\draw[->,blue] (A-1-1.east) to (A-1-2.west);
\draw[->,blue] (A-1-2.east) to (A-1-3.west);
\draw[->,blue] (A-1-3.east) to (A-1-4.west);
\draw[->,blue] (A-1-4.south west) to (A-2-1.north east);
\draw[->,blue] (A-2-1.east) to (A-2-2.west);
\draw[->,blue] (A-2-2.east) to (A-2-3.west);
\draw[->,blue] (A-2-3.east) to (A-2-4.west);
\draw[->,blue] (A-2-4.south west) to (A-3-1.north east);
\draw[->,blue] (A-3-1.east) to (A-3-2.west);
\draw[->,blue] (A-3-2.east) to (A-3-3.west);
\draw[->,blue] (A-3-3.east) to (A-3-4.west);
\draw[->,blue] (A-3-4.south west) to (A-4-1.north east);
\draw[->,blue] (A-4-1.east) to (A-4-2.west);
\draw[->,blue] (A-4-2.east) to (A-4-3.west);
\draw[->,blue] (A-4-3.east) to (A-4-4.west);
\end{tikzpicture}%
}%
}_{\text{Row Major Order}}
\quad 
\underbrace{%
\mbox{%
\begin{tikzpicture}[ampersand replacement=\&]
\matrix (A) [matrix of math nodes,inner sep=0pt,left delimiter={[},right delimiter={]},column sep={8pt}, row sep={8pt}] 
{ 0 \& 4 \& 8 \& 12 \\
  1 \& 5 \& 9 \& 13 \\
  2 \& 6 \& 10 \& 14 \\
  3 \& 7 \& 11 \& 15 \\
};
\draw[->,blue] (A-1-1.south) to (A-2-1.north);
\draw[->,blue] (A-2-1.south) to (A-3-1.north);
\draw[->,blue] (A-3-1.south) to (A-4-1.north);
\draw[->,blue] (A-4-1.north east) to (A-1-2.south west);
\draw[->,blue] (A-1-2.south) to (A-2-2.north);
\draw[->,blue] (A-2-2.south) to (A-3-2.north);
\draw[->,blue] (A-3-2.south) to (A-4-2.north);
\draw[->,blue] (A-4-2.north east) to (A-1-3.south west);
\draw[->,blue] (A-1-3.south) to (A-2-3.north);
\draw[->,blue] (A-2-3.south) to (A-3-3.north);
\draw[->,blue] (A-3-3.south) to (A-4-3.north);
\draw[->,blue] (A-4-3.north east) to (A-1-4.south west);
\draw[->,blue] (A-1-4.south) to (A-2-4.north);
\draw[->,blue] (A-2-4.south) to (A-3-4.north);
\draw[->,blue] (A-3-4.south) to (A-4-4.north);
\end{tikzpicture}%
}%
}_{\text{Column Major Order}}
\qquad 
\underbrace{%
\mbox{%
\begin{tikzpicture}[ampersand replacement=\&]
\matrix (A) [matrix of math nodes,inner sep=0pt,left delimiter={[},right delimiter={]},column sep={8pt}, row sep={8pt}] 
{ 0 \& 1 \& 4 \& 5 \\
  2 \& 3 \& 6 \& 7 \\
  8 \& 9 \& 12 \& 13 \\
 10 \& 11 \& 14 \& 15 \\
};
\draw[->,blue] (A-1-1.east) to (A-1-2.west);
\draw[->,blue] (A-1-2.south west) to (A-2-1.north east);
\draw[->,blue] (A-2-1.east) to (A-2-2.west);
\draw[->,blue] (A-2-2.north east) to (A-1-3.south west);
\draw[->,blue] (A-1-3.east) to (A-1-4.west);
\draw[->,blue] (A-1-4.south west) to (A-2-3.north east);
\draw[->,blue] (A-2-3.east) to (A-2-4.west);
\draw[->,blue] (A-2-4.south west) to (A-3-1.north east);
\draw[->,blue] (A-3-1.east) to (A-3-2.west);
\draw[->,blue] (A-3-2.south west) to (A-4-1.north east);
\draw[->,blue] (A-4-1.east) to (A-4-2.west);
\draw[->,blue] (A-4-2.north east) to  (A-3-3.south west);
\draw[->,blue] (A-3-3.east) to (A-3-4.west);
\draw[->,blue] (A-3-4.south west) to (A-4-3.north east);
\draw[->,blue] (A-4-3.east) to (A-4-4.west);
\end{tikzpicture}%
}%
}_{\text{Row Morton Order}}
\quad
\underbrace{%
\mbox{%
\begin{tikzpicture}[ampersand replacement=\&]
\matrix (A) [matrix of math nodes,inner sep=0pt,left delimiter={[},right delimiter={]},column sep={8pt}, row sep={8pt}] 
{  0 \& 2 \& 8 \& 10 \\
 1 \& 3 \& 9 \& 11 \\
 4 \& 6 \& 12 \& 14 \\
 5 \& 7 \& 13 \& 15 \\
};
\draw[->,blue] (A-1-1.south) to (A-2-1.north);
\draw[->,blue] (A-2-1.north east) to (A-1-2.south west);
\draw[->,blue] (A-1-2.south) to (A-2-2.north);
\draw[->,blue] (A-2-2.south west) to (A-3-1.north east);
\draw[->,blue] (A-3-1.south) to (A-4-1.north);
\draw[->,blue] (A-4-1.north east) to (A-3-2.south west);
\draw[->,blue] (A-3-2.south) to (A-4-2.north);
\draw[->,blue] (A-4-2.north east) to (A-1-3.south west);
\draw[->,blue] (A-1-3.south) to (A-2-3.north);
\draw[->,blue] (A-2-3.north east) to (A-1-4.south west);
\draw[->,blue] (A-1-4.south) to (A-2-4.north);
\draw[->,blue] (A-2-4.south west) to (A-3-3.north east);
\draw[->,blue] (A-3-3.south) to (A-4-3.north);
\draw[->,blue] (A-4-3.north east) to (A-3-4.south west);
\draw[->,blue] (A-3-4.south) to (A-4-4.north);
\end{tikzpicture}%
}%
}_{\text{Column Morton Order}}
\end{equation*}
The name ``Z-curve'' comes from the description of the row Morton order, which enumerates the first $2$-by-$2$ block in a ``Z'' shape, and then recursively repeats that in growing $2$-by-$2$ blocks. For the general Morton code, notice the recursive pattern as well. We are going to use \emph{column} Morton orders in this paper.

More specifically, a Morton code refers to the method used to generate the linear index from the row and column, or the inverse map, which takes a linear Morton index and produces the row and column in the matrix
\begin{equation}
\begin{aligned}
\text{MortonEncode}(i,j) & \to I \in [0, n^2-1] \\
\text{MortonDecode}(I) & \to (i,j) \in [0,n-1]^2.
\end{aligned} 
\end{equation} 
At this point, we've fully switched to programmatic indices as otherwise the confusion between the codes and the text would become extreme. Matrices and vectors, now, are indexed from $0$ as well. 
In the previous $4$-by-$4$ example of the orders, $\text{MortonEncode}(1,2) = 9$, and $\text{MortonDecode}(7) = (3,1)$. (Our row and column indices start at zero if these numbers seem off by one!) The encoding and decoding procedures work on bit-strings for the respective numbers. In fact, the procedure is surprisingly elegant: take the bit-strings for the row and column index and interleave the digits. For example, let's say we want to encode the numbers 1 and 2 into a single value. First, we convert both numbers into base 2: $1 \to 01_2$ and $2 \to 10_2$. Next we interleave the digits starting from the column:
\begin{center}
\begin{tikzpicture}[xscale=0.5,yscale=0.75]
\coordinate (A) at (0,0);
\coordinate (B) at (1,0);
\coordinate (C) at (3,0);
\coordinate (D) at (4,0); 
\coordinate (E) at (0.5,-1);
\coordinate (F) at (1.5,-1);
\coordinate (G) at (2.5,-1);
\coordinate (H) at (3.5,-1);
\draw [-,blue] (A) -- (E);
\draw [-,blue] (B) -- (G);
\draw [-,blue] (C) -- (F);
\draw [-,blue] (D) -- (H);
\node [fill=white,inner sep=2pt] at (A) {1};
\node [fill=white,inner sep=2pt] at (B) {0};
\node [fill=white,inner sep=2pt] at (C) {0};
\node [fill=white,inner sep=2pt] at (D) {1};
\node [fill=white,inner sep=2pt] at (E) {1};
\node [fill=white,inner sep=2pt] at (F) {0};
\node [fill=white,inner sep=2pt] at (G) {0};
\node [fill=white,inner sep=2pt] at (H) {1};
\end{tikzpicture}
\end{center}
The resulting code $1001_2 = 9$: the desired Morton index.
To decode a Morton index into the row and column numbers, we simply reverse the process. We consider $7$ with it's base-2 representation: $0111_2$ and divide the digits into two groups, alternating every binary digit:
\begin{center}
\begin{tikzpicture}[xscale=0.5,yscale=0.75]
\coordinate (A) at (0,0);
\coordinate (B) at (1,0);
\coordinate (C) at (3,0);
\coordinate (D) at (4,0); 
\coordinate (E) at (0.5,1);
\coordinate (F) at (1.5,1);
\coordinate (G) at (2.5,1);
\coordinate (H) at (3.5,1);
\draw [-,blue] (A) -- (E);
\draw [-,blue] (B) -- (G);
\draw [-,blue] (C) -- (F);
\draw [-,blue] (D) -- (H);
\node [fill=white,inner sep=2pt] at (A) {0};
\node [fill=white,inner sep=2pt] at (B) {1};
\node [fill=white,inner sep=2pt] at (C) {1};
\node [fill=white,inner sep=2pt] at (D) {1};
\node [fill=white,inner sep=2pt] at (E) {0};
\node [fill=white,inner sep=2pt] at (F) {1};
\node [fill=white,inner sep=2pt] at (G) {1};
\node [fill=white,inner sep=2pt] at (H) {1};
\end{tikzpicture}
\end{center}
This gives a column index of $1$ and a row index of $3$.

\begin{figure}[t]
\centering
\includegraphics[width=.7\textwidth]{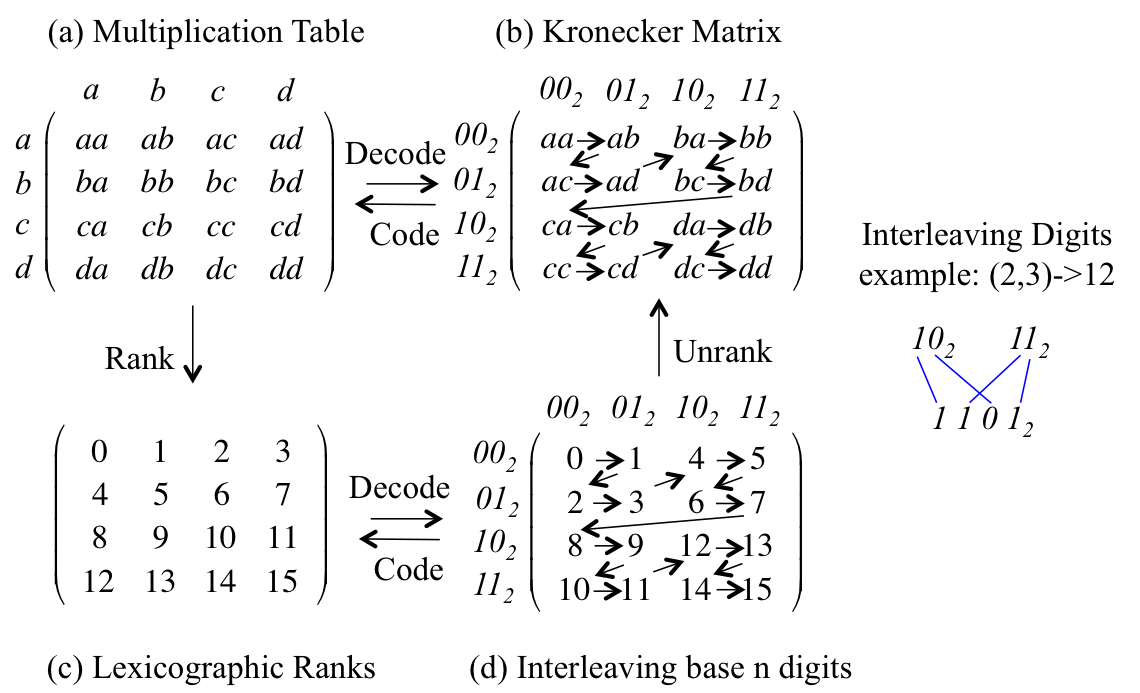}
\caption{The Morton decoding process of a $2^{nd}$ power Kronecker graph. The decoding process takes the lexicographic index of a probability sequence and maps it to 2-dimensional $(i,j)$ indices by a z-order curve. In (a) and (c), the probability sequences are indexed directly in lexicographic order. Writing the index in base $2$ and separating every other digit yields the row and column indices for each permutation as seen in (b). The placing of permutations in the matrix that arises is the same as that of the $2^{nd}$ power Kronecker matrix (d). The ordering of the lexicographic indexes in (b) is a z-shaped fractal pattern (Follow the ordering of the numbers from $0-15$.) }
\label{fig:morton-small}
\end{figure}

Now, suppose we consider the Morton code as map between two linear indices: the column Morton order and the column major order. This is easy to build if we are given a MortonDecode function because we can quickly move from the row and column values to the linear index in column major order. (In case this isn't clear, given the row $r$ and column $c$ indices, the column major index is $r + cn$ where $n$ is the number of rows of the matrix.) The result is a permutation $[0,n^2-1] \to [0,n^2-1]$: 
\begin{equation}
\begin{array}{cccc}
0 \to 0 & 4 \to 2 & 8 \to 8 & 12 \to 10 \\
1 \to 1 & 5 \to 3 & 9 \to 9 & 13 \to 11\\
2 \to 4 & 6 \to 6 & 10 \to 12 & 14 \to 14 \\
3 \to 5 & 7 \to 7 & 11 \to 13 & 15 \to 15
\end{array}.
\end{equation}
This permutation induces a permutation matrix $\mM$:
\begin{equation}
\sbmat{
1 & 0 & 0 & 0 & 0 & 0 & 0 & 0 & 0 & 0 & 0 & 0 & 0 & 0 & 0 & 0 \\
0 & 1 & 0 & 0 & 0 & 0 & 0 & 0 & 0 & 0 & 0 & 0 & 0 & 0 & 0 & 0 \\
0 & 0 & 0 & 0 & 1 & 0 & 0 & 0 & 0 & 0 & 0 & 0 & 0 & 0 & 0 & 0 \\
0 & 0 & 0 & 0 & 0 & 1 & 0 & 0 & 0 & 0 & 0 & 0 & 0 & 0 & 0 & 0 \\
0 & 0 & 1 & 0 & 0 & 0 & 0 & 0 & 0 & 0 & 0 & 0 & 0 & 0 & 0 & 0 \\
0 & 0 & 0 & 1 & 0 & 0 & 0 & 0 & 0 & 0 & 0 & 0 & 0 & 0 & 0 & 0 \\
0 & 0 & 0 & 0 & 0 & 0 & 1 & 0 & 0 & 0 & 0 & 0 & 0 & 0 & 0 & 0 \\
0 & 0 & 0 & 0 & 0 & 0 & 0 & 1 & 0 & 0 & 0 & 0 & 0 & 0 & 0 & 0 \\
0 & 0 & 0 & 0 & 0 & 0 & 0 & 0 & 1 & 0 & 0 & 0 & 0 & 0 & 0 & 0 \\
0 & 0 & 0 & 0 & 0 & 0 & 0 & 0 & 0 & 1 & 0 & 0 & 0 & 0 & 0 & 0 \\
0 & 0 & 0 & 0 & 0 & 0 & 0 & 0 & 0 & 0 & 0 & 0 & 1 & 0 & 0 & 0 \\
0 & 0 & 0 & 0 & 0 & 0 & 0 & 0 & 0 & 0 & 0 & 0 & 0 & 1 & 0 & 0 \\
0 & 0 & 0 & 0 & 0 & 0 & 0 & 0 & 0 & 0 & 1 & 0 & 0 & 0 & 0 & 0 \\
0 & 0 & 0 & 0 & 0 & 0 & 0 & 0 & 0 & 0 & 0 & 1 & 0 & 0 & 0 & 0 \\
0 & 0 & 0 & 0 & 0 & 0 & 0 & 0 & 0 & 0 & 0 & 0 & 0 & 0 & 1 & 0 \\
0 & 0 & 0 & 0 & 0 & 0 & 0 & 0 & 0 & 0 & 0 & 0 & 0 & 0 & 0 & 1}
\end{equation}
The essence of our final primitive is that this Morton map translates between the multiplication table and the Kronecker product as illustrated in Figure~\ref{fig:morton-small}. In this case, we have validated the case of equation~\eqref{eq:morton-kron}
\begin{equation}
\underbrace{\text{vec}(\mK \kron \mK)}_{\substack{\text{Kronecker matrix} \\ \text{in column-major}}}
\quad = \quad \underbrace{\vphantom{()}\mM}_{\substack{\text{MortonDecode} \\ \text{to column-major}}} \quad  \underbrace{[ \tvec(\mK) \kron \tvec(\mK) ] }_{\substack{\text{linear index} \\ \text{from multiplication table}}} 
\end{equation}
where $\mK$ is two-by-two. We prove this statement in full generality shortly. 

\begin{figure}[ht]
\centering
\includegraphics[width=\textwidth]{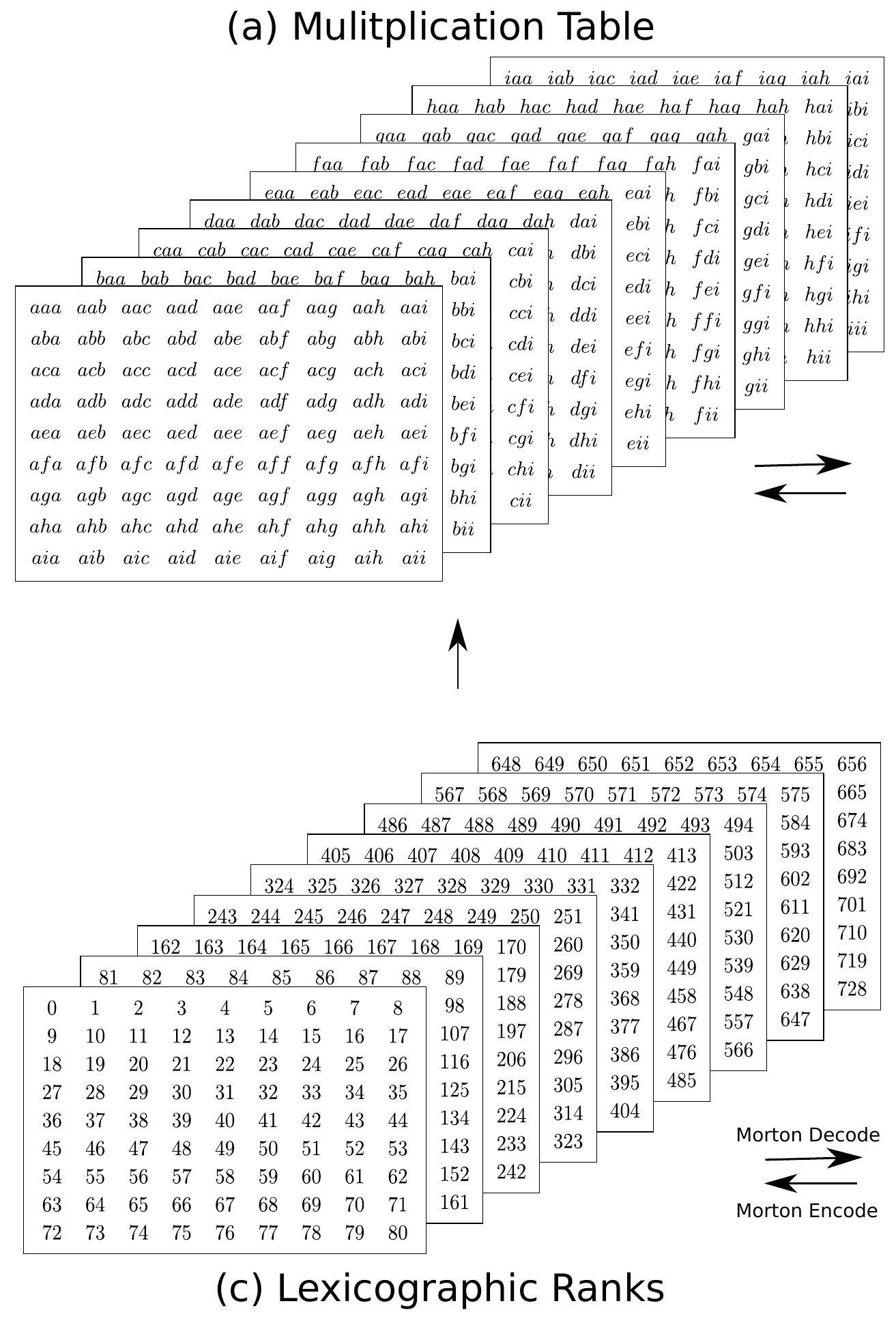}
\caption{The connection between the multiplication table and a repeated Kronecker
matrix with a $3 \times 3$ matrix $\mK$ with $k = 3$. This gives a 9 element vector 
$\vv$ and a $9 \times 9 \times 9$ multiplication table for a $27 \times 27$ matrix
of probabilities. }
\label{fig:morton-big}
\end{figure}

More generally, Morton codes can be defined with respect to any base. We illustrate a case of equation~\eqref{eq:morton-kron} in Figure~\ref{fig:morton-big} where $\mK$ is $3 \times 3$ and $k = 3$. Suppose, in the multiplication table, we generate an edge at index $329$. 
The value $329$ in base $3$ is $110012_3$. The resulting MortonDecode is 
\begin{center}
\begin{tikzpicture}[xscale=0.5,yscale=0.75]
\coordinate (I1) at (0,0);
\coordinate (I2) at (1,0);
\coordinate (I3) at (2,0);
\coordinate (I4) at (3,0); 
\coordinate (I5) at (4,0); 
\coordinate (I6) at (5,0); 
\coordinate (C1) at (-0.5,-1);
\coordinate (C2) at (0.5,-1);
\coordinate (C3) at (1.5,-1);
\coordinate (R1) at (3.5,-1);
\coordinate (R2) at (4.5,-1);
\coordinate (R3) at (5.5,-1);
\draw [-,blue] (I1) -- (C1);
\draw [-,blue] (I2) -- (R1);
\draw [-,blue] (I3) -- (C2);
\draw [-,blue] (I4) -- (R2);
\draw [-,blue] (I5) -- (C3);
\draw [-,blue] (I6) -- (R3);
\node [fill=white,inner sep=2pt] at (I1) {1};
\node [fill=white,inner sep=2pt] at (I2) {1};
\node [fill=white,inner sep=2pt] at (I3) {0};
\node [fill=white,inner sep=2pt] at (I4) {0};
\node [fill=white,inner sep=2pt] at (I5) {1};
\node [fill=white,inner sep=2pt] at (I6) {2};
\node [fill=white,inner sep=2pt] at (C1) {1};
\node [fill=white,inner sep=2pt] at (C2) {0};
\node [fill=white,inner sep=2pt] at (C3) {1};
\node [fill=white,inner sep=2pt] at (R1) {1};
\node [fill=white,inner sep=2pt] at (R2) {0};
\node [fill=white,inner sep=2pt] at (R3) {2};
\end{tikzpicture}
\end{center}
which gives a column index of $1\cdot 3^2 + 1=10$ and a row-index of $1\cdot 3^2 + 2 = 11$. Looking up this entry in the Kronecker matrix (Figure 8, upper right), yields a value of $eaf$. Checking back in the multiplication table (Figure 8, upper left) shows that entry $329$ corresponds to entry $eaf$ as well! (Note that given the base 10 value $329$ we can determine its value in the multiplication table by doing a base $9$ decode. Indeed, $329_{10} = 405_{9}$, which gives a multiset index of $[4,0,5]$ corresponding to the values $v(5) = e, v(1) = a, v(6) = f$.)

Returning to the problem of generating a random Kronecker graph, recall that equation~\eqref{eq:morton-kron} gave us a way of mapping from elements of the multiplication table to elements of the Kronecker matrix. This relationship, then, provides a computational tool to determine where an arbitrary element of the multiplication table generated by \texttt{grass\_hop\_region} occurs in the Kronecker matrix itself, providing the final piece of the fast Kronecker sampling methodology. 

The code that maps indices of the multiplication table through Morton codes to the Kronecker indices is given in Listing~\ref{code:mult_to_kron}. This code assumes that equation~\eqref{eq:morton-kron} is correct, which we will prove in the next section.  In the code, the index in the multiplication table is converted into a single linear index. This conversion is just a base $n^2$ to base 10 conversion. Next, the linear index is MortonDecoded in base $n$. The decoding proceeds by assigning the least significant digit to the row index, then the next digit to the column, the next digit to the row, and so on. The most-significant digit then goes to the column value. 

\begin{listing}
\caption{Mapping multiplication table indices to row and column indices of the Kronecker matrix}
\label{code:mult_to_kron}
\begin{multicols}{2}
\begin{minted}[fontsize=\footnotesize]{python}
""" Map a multi-index from the mult. table 
table to a row and column in the Kronecker 
matrix. The input is:
  mind: the multi-index for the mult table
  n: the size of the initiator matrix K
Example:
  map_mult_to_kron([1,3],2)   # = (3,1)
  map_mult_to_kron([4,0,7],3) # = (10,11)
"""
def map_mult_to_kron(mind,n):
  I = multiindex_to_linear(mind,n*n)
  return morton_decode(I,n)
  
def multiindex_to_linear(mind,n2):
  I = 0
  base = 1
  for i in xrange(len(mind)-1,-1,-1):
    I += mind[i]*base
    base *= n2
  return I
  
  
def morton_decode(I,n):
  row = 0
  rowbase = 1
  col = 0
  colbase = 1
  i = 0
  while I > 0:
    digit = I%n 
    I = I // n
    if i%2 == 0:
      row += rowbase*digit
      rowbase *= n
    else:
      col += colbase*digit
      colbase *= n
    i += 1
  return (row,col)
\end{minted}
\end{multicols}
\end{listing}

What remains is the proof that equation~\eqref{eq:morton-kron} is correct,
which we tackle in the next section.

\newpage
\newpage

\begin{figure}[ht]
\centering
\includegraphics[width=\textwidth]{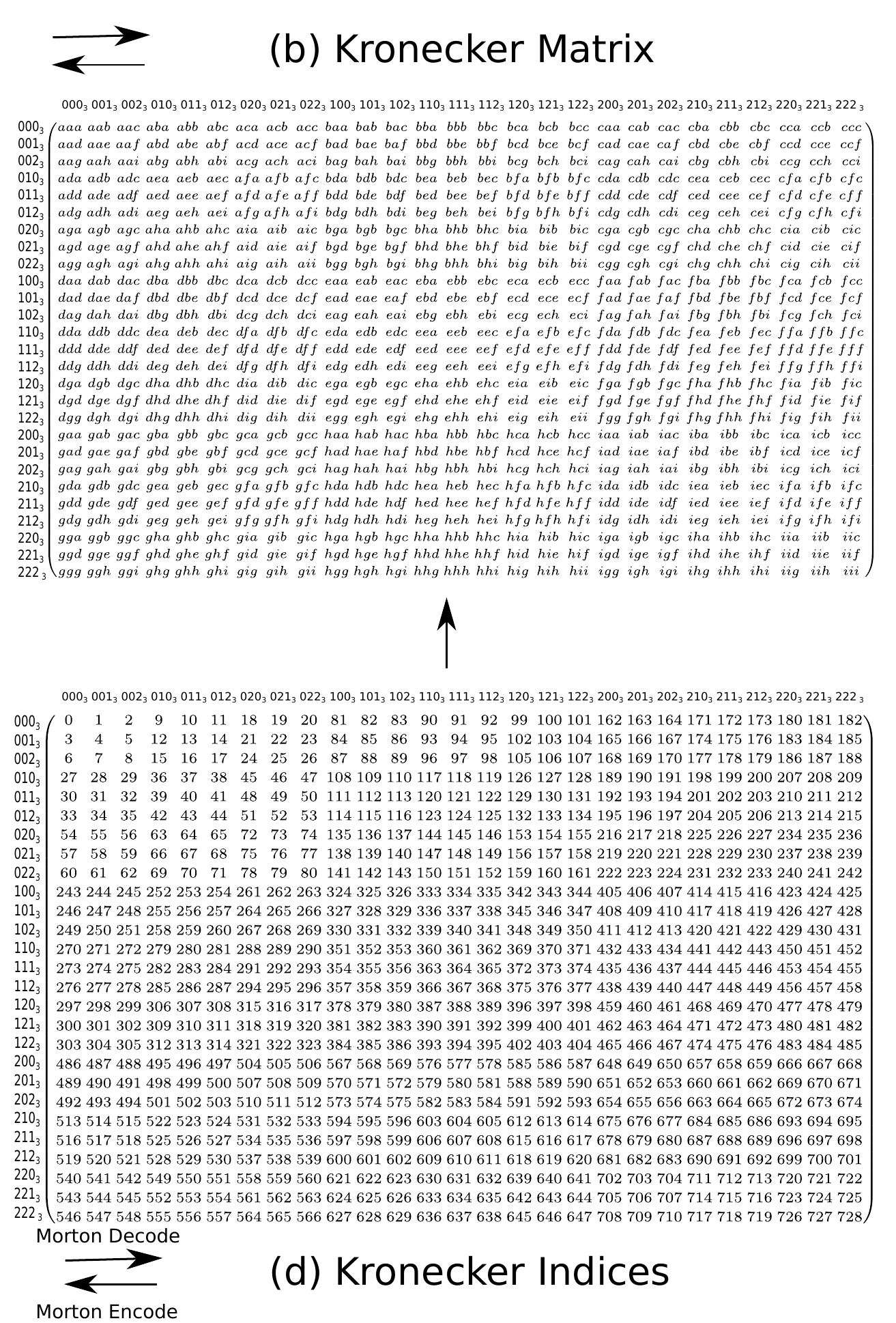}
\footnotesize
    \vskip\abovecaptionskip
    {\normalfont\scshape Fig. \ref{fig:morton-big}}. {\normalfont\itshape (continued)}\par
    
    \vskip\belowcaptionskip
\label{xxxx2}
\end{figure}
\clearpage

\subsection{The proof of the Multiplication Table to Kronecker matrix permutation}

In this section, we will prove \eqref{eq:morton-kron} via the following theorem.

\begin{theorem} 
\label{thm:morton}
Let $\mK$ be a $n \times n$ matrix and $\vv = \tvec(\mK)$ be the column-major representation of $\mK$. Consider an element in the multiplication table 
\[ \underbrace{\vv \kron \vv \kron \cdots \kron \vv}_{k \text{ terms }} \] 
with index $(r_1, r_2, \ldots, r_k)$. Let $I$ be the lexicographic index of the element $(r_1, \ldots, r_k)$. Then the base $n$ Morton decoding of $I$ provides the row and column indices of an element in 
\[ \underbrace{\mK \kron \mK \kron \cdots \kron \mK}_{k \text{ terms}} \] 
with the same value. 
If we convert the row and column indices provided by Morton into a column-major index, then we can build a permutation matrix $\mM_{n,k}$ in order to write: 
\[ \tvec(\underbrace{\mK \kron \mK \kron \cdots \kron \mK}_{k \text{ terms}}) = \mM_{n,k} ( \underbrace{\vv \kron \vv \kron \cdots \kron \vv}_{k \text{ terms }} ). \]
\label{thm}
\end{theorem}

\begin{proof}
To state the proof concisely, let $\mK^{\kron k} = \mKk$, and $\vv^{\kron k}$ be the same. In the proof, we will just show that the Morton code for $I$ gives the correct row and column indices for $\mK^{\kron k}$. The resulting permutation matrix follows by converting those row and column indices to a column-major index and building the full permutation matrix. We will prove this statement by induction on the power of the Kronecker product, $k$.

\emph{Base case.} When $k = 1$, $\mK^{\kron k} = \mK$ is simply the initiator matrix. Consequently, the theorem reads $\tvec(\mK) = \mM_{n,1} \vv$. But the length $1$ Morton code is simply the row and column index in column major order, and so this result follows because the map $\mM_{n,1}$ is the identity. 

\emph{Inductive step overview.}
We will prove our statement inductively by calculating the new row and column values in two ways: first by recursively defining the row and column values using a ``zooming'' argument, and second by applying the Morton decoding process on a recursively defined lexicographic index, $I_{k+1}$. The following figure shows an illustrative overview of how we will show the result in general. 
\begin{center}
\resizebox{.9\linewidth}{!}{%
\begin{tikzpicture}[yscale=0.75]
\draw [gray] (-1,0) rectangle (4,2) node[pos=.5,black] {Multiplication Table Index};
\draw [gray] (6,2) rectangle (10,4) node[pos=.5,black] {Lexicographical Index $I$};
\draw [gray] (12,0) rectangle (17,2) node[pos=.5,black] {$(i,j)$ row and  column location};
\draw[<->, ultra thick] (2.4,2.2) -- (5.8,2.8) node [midway, above, sloped] (TextNode) {base $N^2$ to base $10$};
\draw[->, ultra thick] (4.2,1) -- (11.8,1) node [midway, below, sloped] (TextNode) {`Zooming', Backward-Mapping};
\draw[<->, ultra thick] (10.2,2.8) -- (13.6,2.2) node [midway, above, sloped] (TextNode) {MortonDecode};
\end{tikzpicture}
}
\end{center}

\emph{Inductive step formal.} 
Assume $\mM_{n,k}$ correctly maps between $\vv^{\kron k}$ and $\mK^{\kron k}$. More specifically, this means we can find the row and column index $(R_k, C_k)$ for a given index of the multiplication table $(r_1, r_2, \ldots, r_k)$.

Let $(r_1, r_2, \ldots, r_k, r_{k+1})$ be the multiplication table index for an entry of $\vv^{\kron k+1}$ Then notice that $(r_1, r_2, \ldots, r_k)$ corresponds to an index in $\vv^{\kron k}$. By assumption, the Morton code correctly gives the correct row and column values in $\mK^{\kron k}$: $R_k$ and $C_k$.
Note that the structure of the $\mK^{\kron k+1}$ is: 
\begin{equation}
\left[ \begin{array}{c|c|c|c}
K_{1,1} \mK^{\kron k} & K_{1,2} \mK^{\kron k}& \cdots & K_{1,n} \mK^{\kron k} \\ \hline
K_{2,1} \mK^{\kron k} & K_{2,2} \mK^{\kron k}& \cdots & K_{2,1} \mK^{\kron k}\\  \hline
\vdots & \vdots & \ddots & \vdots \\  \hline
K_{n,1} \mK^{\kron k} & K_{n,2} \mK^{\kron k} & \cdots & K_{n,n} \mK^{\kron k} 
\end{array} \right]
\end{equation}
Then the row value for $(r_1, r_2, \ldots, r_k, r_{k+1})$
\begin{equation} 
R_{k+1} = R_k +  (r_{k+1} \bmod n)n^k.
\label{eqn:rowval}
\end{equation}
The reasoning for this is as follows. The size of the previous iteration's matrix is $n^k$. When we incorporate the new value $r_{k+1}$, we move to one of the $n^2$ areas (of dimension $n^k$ by $n^k$) of the Kronecker matrix where $(r_1, r_2, \ldots, r_k)$ is the suffix. The specific row of the region can be found by simply taking the last multiplication table index $r_{k+1}$ mod $n$, which corresponds to looking up the row index for $r_{k+1}$ in column-major order.
The same reasoning can be translated to find the column index, $j_{k+1}$. 
\begin{equation}
C_{k+1} = C_k + \left\lfloor {\frac{r_{k+1}}{n}} \right\rfloor n^k
\label{eqn:colval}
\end{equation}
By taking the floor of $r_{k+1}$ divided by $n$, we know the column index of the value for $r_{k+1}$. We call this procedure \emph{backward-mapping} or \emph{zooming} because we use the inductive step to \emph{zoom} into the right cell. 


Now, we must show that the Morton decoding process arrives at the same updated row and column values. 
First we will calculate the lexicographic index of our length $k+1$ multiplication table index. A given multiplication table index $(r_1, r_2, \ldots, r_k)$ has a lexicographic index $I_k$. When we move to the next Kronecker power, we incorporate the value $r_{k+1}$. We can determine the new lexicographic index $I_{k+1}$ using $I_k$ and $r_{k+1}$. 
\begin{equation}
I_{k+1} = I_k + r_{k+1}{(n^2)}^k
\label{eqn:rank}
\end{equation}
This relationship arises because the lexicographic index can be calculated by converting our multiplication table index from base $n^2$ to base 10. The last value $r_{k+1}$ will occupy the place value at the front of linearized index, i.e. ${(n^2)}^k$.

Now we must show that Morton decoding correctly maps $I_{k+1}$ to $R_{k+1}$ and $C_{k+1}$. Using the division and remainder theorem on $r_{k+1}$ divided by $n$ in equation \eqref{eqn:rank} results in
\begin{equation}
\label{eqn:remainder}
\begin{aligned} 
r_{k+1} = \lfloor {\frac{r_{k+1}}{n}} \rfloor n + (r_{k+1} \bmod n) 
\end{aligned}
\end{equation}
Substituting this representation into \eqref{eqn:rank} yields
\begin{equation} 
I_{k+1} = \lfloor {\frac{r_{k+1}}{n}} \rfloor n^{2k+1} + (r_{k+1} \bmod n)n^{2k} + I_k
\label{eqn:subsituted}
\end{equation}
Thus, the last two digits that Morton decoding will find here are $\lfloor {\frac{r_{k+1}}{n}} \rfloor$ and $(r_{k+1} \bmod n)$, because these correspond to the final terms in a base $n$ representation of $I_{k+1}$. Consequently, Morton decoding of $I_{k+1}$ is equivalent to decoding $I_k$ and appending the digits $(r_{k+1} \bmod n)$ and $\lfloor {\frac{r_{k+1}}{n}} \rfloor$ to the beginning of the new row and column digit-strings respectively.
This completes our proof because equation \eqref{eqn:subsituted} gives precisely the row and column values found in equations \eqref{eqn:rowval} and \eqref{eqn:colval}.
\end{proof}

Note that this proof actually suggests a more straightforward way to generate the row and column indices of the Kronecker matrix entries, see Problem~\ref{prob:backwards} for more information. 
\section{Problems and Pointers}

We completed our major goal, which was to establish a new link between Morton codes and repeated Kronecker products and described how that result is useful in efficiently generating a Kronecker graph. While we have not focused on producing the most efficient implementation possible (see Problem~\ref{prob:backwards} for a non-trivial improvement), we did focus on a procedure that only does a small amount of additional work per edge and is faster than ball-dropping procedures that have dominated the experimental landscape. 

Our work builds strongly upon the work of~\citet{moreno}, and highlights new connections to classic problems in combinatorics including ranking and unranking~\cite{bonet}. Indeed, the ranking and unranking perspective has a tremendous amount to offer for ever more complicated models of graphs. For instance, the original specification of the \ERfull graph due to~\citet{Erdos-1959-random} was a random choice of graphs with $n$ vertices and $m$ edges. It is straightforward to sample such an object in the ranking and unranking perspective. To each graph, we associate a binary string of length $\binom{n}{2}$, which represents the strictly upper-triangular region of an adjacency matrix. This binary string must have exactly $m$ ones. To randomly generate such a string, first generate a random number between $0$ and $\binom{\binom{n}{2}}{m}-1$, then \emph{unrank} that random number in lexicographic order. This procedure follows by testing if the first number is $0$ or $1$. There are $\binom{\binom{n}{2} - 1}{m}$ choices where the leading value is a $0$ and $\binom{\binom{n}{2} - 1}{m-1}$ choices where it is a $1$, so we test if our number is larger than $\binom{\binom{n}{2} - 1}{m}$ and then assign the first digit. This process can be repeated to deterministically generate the string. This process could then be optimized by incorporating strategies to improve the speed and efficiency~\cite{bonet}. 



We conclude our manuscript by (i) reviewing the major timeline of the literature, (ii) pointing out some strongly related work that falls outside of our specific scope, and (iii) providing some follow-up problems. 

\subsection{Recap of major literature}
\begin{compactitem}
\item 1959: \ERfull paper~\cite{Erdos-1959-random} with $m$ fixed edges
\item 1959: Gilbert's model for $G(n,p)$~\cite{Gilbert-1959-random} 
\item 1962: Paper on grass-hopping or leap-frogging for sequential trials in statistics~\cite{FanMullerRezucha1962}
\item 1983: Stochastic block model proposed~\cite{Holland-1983-sbm}
\item 2002: Chung-Lu graphs proposed~\cite{Chung-2002-random}
\item 2004: RMAT, Kronecker predecessor, proposed~\cite{Chakrabarti-2004-rmat}
\item 2005: Kronecker graphs proposed~\cite{leskovec2005}
\item 2005: Grass-hopping for \ERfull graphs~\cite{BatageljBrandes2005}
\item 2011: Observation on \ERfull blocks in Chung-Lu~\cite{MillerHagberg2011}
\item 2013: Observation on \ERfull blocks in Kronecker~\cite{moreno}
\end{compactitem}

\subsection{Pointers to more complex graph models}

The models we consider here are not considered state of the art, although they are often used as synthetic examples due to their simplicity and ubiquity. In this section, we provide a few references to more intricate models that could be considered state of the art.  We've attempted to limit our discussion to the most strongly related literature. A full survey of random graph models is well beyond our scope. 

\begin{description}
\item[{\footnotesize{\textsc{BTER}}},~\citet{Kolda-2014-BTER}] The block-two-level \ERfull ({\footnotesize{\textsc{BTER}}}) graph models a collection of \ERfull subgraphs akin to the diagonal blocks of a stochastic block models. However, it also includes a degree constraint that better models real-world networks. 
\item[m{\footnotesize{\textsc{KPGM}}},~\citet{Moreno-2010-mKPMG}] Another model of interest is called the mixed Kronecker Product Graph Model (m{\footnotesize{\textsc{KPGM}}}), which is a generalization of the Kronecker model, but allows for more variance that you may expect in real world populations of networks. The main difference in this model from the standard Kronecker model that we explored, is that it is based on realizing an adjacency matrix at stages of the Kronecker product calculation. This corresponds to increasing or zeroing probabilities in the final Kronecker product, e.g. $\mP = \mA_\ell \kron (\mK \kron \cdots \text{ $k-\ell$ terms } \cdots \kron \mK)$ where $\mA_{\ell}$ is a realization of an $\ell$-level Kronecker graph with $\mK$. 
\item[Random Kernel Graph,~\citet{RSA-2007-inhomogenous-random-graphs}] This models a general, but specific, structure for $\mP$, where $P_{i,j} = \min(f(v_i, v_j)/n,1)$ and $f$ is a kernel function with a few special properties, and $v_i = i/n$ for each vertex.  The Chung-Lu model can be expressed in this form and there is an efficient sampling algorithm due to~\cite{HagbergLemons2015}. 
\item[Multifractal networks,~\citet{Palla2010-multifractal}] These are a continuous generalization of the Kronecker matrix models we consider here and generate a network based on a function $P(x,y)$ where $x$ and $y$ are in $[0,1]$, but also where $P(x,y)$ is the result of repeated convolution of a generator (akin to repeated Kronecker products, see the paper for details). With the resulting measure $P(x,y)$, we draw random coordinates $v_i$ for each node in $[0,1]$ and generate a graph where $P_{ij} = P(v_i, v_j)$. 
\end{description} 
There are strong relationships among these classes of models. We conjecture that this idea of \emph{grass-hopping} can be extended to \emph{all} of these random graph models, and more generally, all random graph generation mechanisms that have a small number of parameters and computationally efficient strategies to compute the probability of a given edge. (For an $n^k$ vertex Kronecker graph, there are $n^2$ parameters and each probability can be computed in $k$ steps.) 

\subsection{Problems}
For the following problems. We will post solutions or solution sketches in the online supplemental materials for our manuscript. However, we encourage readers to solve them individually! 

\begin{problem}[Easy] \label{prob:ball-drop-oversampling}
When we were comparing ball-dropping to grass-hopping for sampling \ERfull graphs, we derived that an approximate count for the number of edges used in ball-dropping is: $m \log(1/(1-p))/p$. First show that the term $ \log(1/(1-p))/p$ has a removable singularity at $p=0$ by showing that $\lim_{p \to 0}  \log(1/(1-p))/p = 1$. Second, show that $ \log(1/(1-p))/p > 1$ if $p > 0$. 
\end{problem}
\begin{solution}
We need to use L'H\^{o}pital's rule for the first question, in which was we get $(1/(1-p)) / 1$ for the ratio of derivatives. This clearly tends to $1$. For the second part, we take the Taylor series of $\log(1/(1-p))$, which is $p + p^2/2 + p^3/3 + \ldots$. This is always at least $p$ when $p > 0$. (Note, that, indeed, this series divided by $p$ gives $1 + p/2 + p^2/3 + \ldots$)
\end{solution}

\begin{problem}[Easy]
\label{prob:chung-lu-grass-hop}
Implement a grass-hopping scheme for Chung-Lu as described in Section~\ref{sec:chung-lu-union}. 
\end{problem}
\begin{solution}
Note that this code uses the method grass$\_$hop$\_$er from listing \ref{code:grass-hop}.
\begin{minted}[fontsize=\footnotesize]{python}
"""Create a Chung-Lu graph via grass-hopping
Note this routine is only for pedagogical use as
It can be improved in many ways.
Example: chung_lu_grasshop([5,1,1,2,2,1])"""
def chung_lu_grasshop(d):
  # sort d with a permutation
  # http://stackoverflow.com/questions/7851077/how-to-return-index-of-a-sorted-list
  dperm = sorted(range(len(d)), key=lambda k: d[k])
  # build regions
  n = 0
  regions = []
  while n < len(dperm):
    first = n
    deg = d[dperm[n]]
    while n < len(dperm)-1 and d[dperm[n+1]] == deg:
      n += 1
    regions.append((first,n))
  # grasshop within regions
  vol = sum(d)
  edges = []
  for ri in xrange(len(regions)):
    for rj in xrange(len(regions)):
      rowinds = regions[ri] # start/end of row regions
      colinds = regions[rj] # start/end of col region
      p = d[dperm[rowinds[0]]]*d[dperm[colinds[0]]]/vol 
      rowsize = rowinds[1] - rowinds[0] + 1
      colsize = colinds[1] - colinds[0] + 1
      er_edges = grass_hop_er(max(rowsize, colsize), p)
      edges.extend( [ (i+rowinds[0], j+colinds[0]) 
      	for i,j in er_edges if i < rowsize and j < colsize ] )
      
  # build inverse permutation to re-permute edge indices
  iperm = [0 for _ in dperm] # inverse permutation
  for i,j in enumerate(dperm):
    iperm[j] = i
  # reverse permute edges
  return [ (iperm[e[0]], iperm[e[1]]) for e in edges ] 
\end{minted}
\end{solution}

\begin{problem}[Easy]
\label{prob:chung-lu-ball-drop}
While we considered coin-flipping, ball-dropping, and grass-hopping for \ERfull, we did not do so for some of the other models. Of these, the Chung-Lu ball-dropping procedure is perhaps the most interesting. Recall how ball-dropping works for \ERfull. We determine how many edges we expect the graph to have. Then we drop balls into random cells until we obtain the desired number of edges. The way we dropped them was so that every single entry of the matrix was equally likely. The procedure for Chung-Lu works in a similar manner, but we won't select each cell equally likely. For simplicity, we describe it using the exact expected number of edges. 

Note that the expected number of edges in a Chung-Lu graph is just 
\[ E[\sum_i \sum_j A_{ij}] =  \sum_i \sum_j E[A_{ij}] =  \sum_i d_i, \]
where we used the result of equation \eqref{eqn:chungexp} in order to introduce $d_i$. 

Let $m = \sum_i d_i$. We will ball-drop until we have $m$ edges. 

The goal of each ball-drop is to pick cell $(i,j)$ with probability $d_i d_j / \sum_{k,l} d_k d_l$. This forms a valid probability distribution over cells. The difference from equation \eqref{eqn:chung-lu} is subtle, but meaningful. Also note that each $d_i$ is an integer. So to sample from this distribution, what we do is build a list where entry $i$ occurs exactly $d_i$ times. If we draw an entry out of this list, then the probability of getting $i$ is exactly $d_i / \sum_{k} d_k$. Your first task is to show that if we draw two entries out of this list, sampling with replacement, the probability of the two drawn entries being $i$ and $j$ is: 
\[ d_i d_j / \sum_{k,l} d_k d_l. \]
This sampling procedure is easy to implement. Just build the list and randomly pick an entry! 

Your problem is then:
\begin{compactitem}
\item show that the probability we computed above is correct
\item implement this procedure on the computer akin to the Python codes we've seen
\item prepare a plot akin to Figure~\ref{fig:balldrop} for the Chung-Lu graphs as you vary the properties of the degree distribution. 
\end{compactitem}
\end{problem} 
\begin{solution}
If we draw two samples from the list. The draws are independent, so the probability of the event is the product of each draw's probability. Each occurs with probability $d_i/\sum_{k} d_k$. So if we saw $i$ and then $j$, this even occurs with probability 
\[ \frac{d_i}{\sum_{k} d_k} \cdot \frac{d_j}{\sum_{k} d_k} = \frac{d_i d_j}{\sum_{k,\ell} d_k d_l}. \]

\begin{minted}[fontsize=\footnotesize]{python}
""" Build a list where entry i occurs d[i] times. """
def build_list(d):
  L = []
  for i in xrange(len(d)):
     L.extend( [i for _ in xrange(d[i])] )
  return L

"""Create a Chung-Lu graph via ball-dropping.
Note this routine is only for pedagogical use as
It can be improved in many ways.
Example: chung_lu_ball_drop([5,1,1,2,2,1])"""
def chung_lu_ball_drop(d):
  M = sum(d)
  L = build_list(d)
  edges = set()
  dups = 0
  while len(edges) < M:
    src = L[random.randint(0,len(L)-1)]
    dst = L[random.randint(0,len(L)-1)]
    if (src,dst) not in edges:
      edges.add((src,dst))
    else:
      dups += 1
  return (list(edges), dups)
\end{minted}
\end{solution}

\begin{problem}[Easy]
In Section~\ref{sec:sbm-union}, we wrote a short code using \verb#grass_hop_er# to generate a two-block stochastic block model (2SBM). In that short code, we generated larger \ERfull blocks for the \emph{off-diagonal blocks} then necessary and only included edges if they fell in a smaller region. This inefficiency, which is severe if $n_1 \gg n_2$ can be addressed for a more general \verb#grass_hop_er# function. 

Implement \verb#grass_hop_er# for an $m \times n$ region and rewrite the short two-block stochastic block model to use this more general routine to avoid the inefficiency 
\end{problem}
\begin{solution}
\begin{minted}[fontsize=\footnotesize]{python}
def grass_hop_er(m,n,p):
    edgeindex = -1                    # we label edges from 0 to n^2-1
    gap = np.random.geometric(p)      # first distance to edge
    edges = []
    while edgeindex+gap < m*n:       # check to make sure we have a valid index
        edgeindex += gap             # increment the index
        src = edgeindex // n         # use integer division, gives src in [0, n-1]
        dst = edgeindex - n*src      # identify the column 
        edges.append((src,dst))
        gap = np.random.geometric(p) # generate the next gap
    return edges
def sbm2(n1,n2,p,q):
  edges = grass_hop_er(n1,n1,p)                       # generate the n1-by-n1 block
  edges.extend( [ (i+n1,j+n1) for i,j in grass_hop_er(n2,n2,p) ])  # n2-by-n2 block
  edges.extend( [ (i,j+n1) for i,j in grass_hop_er(n1,n2,q) ])     # n1-by-n2 block
  edges.extend( [ (i+n1,j) for i,j in grass_hop_er(n2,n1,q) ])     # n2-by-n1 block
  return edges
\end{minted}
\end{solution}

\begin{problem}[Easy]
Suppose that you have the following function implemented: 
\begin{minted}[fontsize=\footnotesize]{python}
""" Generate a general stochastic block model where the input is: 
ns: a list of integers for the size of each block (length k).
 Q: a k-by-k matrix of probabilities for the within block sizes. """
def sbm(ns, Q)
\end{minted}
Describe how you could implement a program to generate a 
Chung-Lu graph that makes a single call to
\verb#sbm#.
\end{problem}
\begin{solution}
Recall from section \ref{sec:chung-lu-union} that we can arrange the nodes of our graph in increasing degree order, which forms a block structure. Sort the nodes in this way, and set $ns = [n_1, \ldots, n_k]$, where $n_i$ is the number of degrees with the $ith$ largest unique degree. Store the values $\frac{d_{[i]}d_{[j]}}{\rho}$ in $Q_{ij}$. Then run the sbm procedure on this output.
\end{solution}

\begin{problem}[Medium, Partially-solved]
\label{prob:sbm-ball-drop}
Implement and evaluate a ball-dropping procedure for the stochastic block model. To determine the number of edges, use the result about the number of edges in an \ERfull block and apply it to each block. To run ball-dropping, first pick a region based on the probability of $\mQ$, (hint, you can do this with binary search and a uniform random variable) then pick within that \ERfull block. 
\end{problem}
\begin{solution}
We are going to sketch a solution and outline some interesting directions to take as far as extending.  The solution involves sampling an entry at random based on  the probabilities in $\mQ$. This can be done with a  routine such as those described here: 
\url{http://stackoverflow.com/questions/4437250/choose-list-variable-given-probability-of-each-variable}. Then, once you have the region of $\mQ$, you ball-drop like in \ERfull. This assumes there is no structure in $\mQ$. A more interesting case is where there is structure in $\mQ$. For instance, where there is a single within group probability $p$ and a single between group probability $q$ such that $\mQ =  (p-q)\mI + q  \ve \ve^T$ or $Q_{ij} =  p$ if $i = j$ and $Q_{ij} = q$ if $i \not= j$, and where all the sizes of each group are equal.  In this case, you can first choose if the ball will go into the diagonal or non-diagonal region. Then conditional on that choice, pick an element of $\mQ$. Then do an \ERfull ball-drop within the region of $\mP$ that corresponds to an element of $\mQ$. 
\end{solution} 

\begin{problem}[Medium]
Write an implementation to generate a stochastic block model graph using 
\verb#grass_hop_er# as a sub-routine. 
\end{problem}

\begin{solution}
\begin{minted}[fontsize=\footnotesize]{python}
#Generate a general stochastic block model using grass-hopping where the input is: 
#    ns: a list of integers for the size of each block (length k).
#    Q: a k-by-k matrix of probabilities for the within block sizes.
def sbm_er(ns, Q):
    edges = [];
    for i in range(len(ns)):
        for j in range(len(ns)): #loop through each block
            m=ns[i]; n=ns[j]; #find m x n size of block
            row_loc=0;
            for k in range(i):
                row_loc+=ns[k]; #locate row of top left corner of block
            col_loc=0;
            for l in range(j):
                col_loc+=ns[l]; #locate column of top left corner of block
            for e in grass_hop_er(m,n,Q[i][j]): #find edges
                edges.append([e[0]+row_loc,e[1]+col_loc]); #add edges updated with top left corner
    return edges;
\end{minted}
\end{solution}

\begin{problem}[Easy]
Extend all the programs in this tutorial to have proper input validation so that they throw easy-to-understand errors if the input is invalid. (For instance, \ERfull graphs need the \verb#p# input to be a probability.)
\end{problem}

\begin{problem}[Medium]
In the paper, we worked out the number of \ERfull regions for a Kronecker graph assuming that the probabilities in $\mK$ were non-symmetric. But in Figure~\ref{fig:kron-regions-small}, we showed the \ERfull regions where the matrix $\mK$ was symmetric. Determine a tight closed form expression for the number of \ERfull regions when $\mK$ is symmetric.
\end{problem}
\begin{solution}
Note that the kth Kronecker power of $\mK$ contains all length-k permutations of the probabilities contained in $\mK$. Thus, it suffices to count the number of unique probabilities in $\mK$ and apply the same stars-and-bars technique shown in \ref{sec:count_regions}. The number of unique values in an n x n symmetric matrix is equivalent to the number of values in the upper-triangular matrix: $\binom{n+1}{2}$. In order to count the number of length-k sequences of these values, we apply the stars-and-bars formula and attain $\binom{\binom{n+1}{2}+k-1}{k}$
\end{solution}


\begin{problem}[Hard]
It turns out that \ERfull's original proposal for graphs and what we now call \ERfull graphs are different! What \citet{Erdos-1959-random} actually proposed was: generate a random adjacency matrix with exactly $m$ non-zeros, where $m$ is specified. This is what we do in the ball-dropping procedure for generating \ERfull graphs, but $m$ is chosen from a binomial distribution to emulate the number of entries that result from coin-flips. 
\end{problem}
\begin{solution}
Test
\end{solution}

\begin{problem}[Easy]
\label{prob:phalf}
Implement a simple-change to \verb#ball_drop_er# that will generate non-edges when $p > 0.5$ and filter the list of all edges. Our own solution to this involves adding four lines. (But we used a few pieces of python syntactic sugar -- the point is that it isn't a complicated idea.) Compare the run times of this approach with the standard ball-dropping approach 
\end{problem}
\begin{solution}
\begin{minted}[fontsize=\footnotesize]{python}
def ball_drop_er(n,p):
  if p > 0.5:
    nonedges = ball_drop_er(n,1-p)
    alledges = set((i,j) for i in xrange(n) for j in xrange(n))
    return list(alledges.difference(nonedges))
  m = int(np.random.binomial(n*n,p)) # the number of edges
  edges = set() # store the set of edges
  while len(edges) < m:
    # the entire ball drop procedure is one line, we use python indices in 0,n-1 here
    e = (random.randint(0,n-1),random.randint(0,n-1))
    if e not in edges: # check for duplicates
       edges.add(e) # add it to the list
  return list(edges) 
\end{minted}
\end{solution}

\begin{problem}[Medium]
\label{prob:stirling's}
In section \ref{sec:count_regions}, we showed that the number of subregions is on the order of $O(k^{n^2-1})$. There are many ways to arrive at this result. Using Stirling's approximation for combinations, find a different way to count the number of subregions. Stirling's Approximation: $\log n! = n\log n - n + \frac{1}{2}\log n + \frac{1}{2}\log (2 \pi) + \epsilon_n$
\end{problem}
\begin{solution}
The number of subregions is given by $\binom{n^2+k-1}{k}$, which expands to $\binom{(n^2+k-1)!}{(n^2-1)!k!}$. Taking the log of our expression allows us separate out the components of the combination and arrive at: $\log (n^2+k-1)!-\log (n^2-1)!-\log(k!)$. Now we can substitute in Stirling's approximation for each $\log k!$ term and simplify arriving at:$k \log \frac{n^2+k-1}{k} +(n^2-1)(\log(n^2+k-1)-1)+\frac{1}{2}\log \frac{n^2+k-1}{k}$. Note that our simplification process is made easier by eliminating all constant terms (those with no k).  Using L'Hopital's Rule, it can be found that $\lim_k\rightarrow \inf k \log \frac{n^2+k-1}{k} =  n^2-1$. Furthermore, at the limit, $\frac{1}{2}\log \frac{n^2+k-1}{k}$ goes to 0. Thus, we can simplify our expression further to $(n^2-1)\log n^2+k-1$. Exponentiating this expression yields $(n^2+k-1)^(n^2-1)$, which is $O(k(n^2-1)$ as desired.
\end{solution}

\begin{problem}[Medium]
\label{prob:backwards}
Note that in the proof of Theorem~\ref{thm:morton}, we began by showing that the multiset region could be easily decoded into the row and column index of the entry in the Kronecker matrix. Adapt our programs to use this insight to avoid computing the linear multiindex index returned by \verb#multiindex_to_linear#.
\end{problem}
\begin{solution}
\begin{minted}[fontsize=\footnotesize]{python}
# backward_map takes as input
#    s: an array (edge probability sequence) consisting of a
#       set letters represented as numbers e.g. [0,1,3,1]
#    n: the size of the n x n initiator matrix 
# and returns the corresponding row or column index of the
# probability sequence in the kronecker product graph matrix
def backward_map(s,n):
    row=0
    col=0
    # we want to "zoom" further into the matrix for each
    # value of our cell
    for i, r in enumerate(s):
        length = len(s)-i;
        row += (r mod n) * math.pow(n,length-1);
        col += int(r/n) * math.pow(n,length-1);
    return [row,col];
\end{minted}
\end{solution}

\begin{problem}[Easy]
When we were unranking multisets, we would compute the number of multisets that started with a digit as the prefix and then enumerate the number of such sets. We used this to identify the first digit of the unranked set. Double-check that we didn't miss any possibilities by summing 
\end{problem}
\begin{solution}
The number of multiset permutations is given by \begin{equation}
\begin{aligned}
\label{eqn:mult}
\frac{k!}{\prod_{i=0}^{n^2} r_i!} \end{aligned}
\end{equation}
Removing an instance of the first letter results in $\frac{(k-1)!r_0}{\prod_{i=0}^{n^2} r_i!}$. In general, we can compute the sum of all permutations with one digit removed as follows: 
\begin{equation}
\begin{aligned}
\label{eqn:mult2}
\sum_{j=1}^{n^2} \frac{(k-1)!r_j}{\prod_{i=0}^{n^2} r_i!}
\end{aligned}
\end{equation}

It suffices to show that \begin{equation}
\begin{aligned}
\frac{k!}{\prod_{i=0}^{n^2} r_i!} = \sum_{j=1}^{n^2} \frac{(k-1)!r_j}{\prod_{i=0}^{n^2} r_i!} \end{aligned}
\end{equation}
Canceling common terms yields \begin{equation}
\begin{aligned}
k= \sum_{j=1}^{n^2} r_j
\end{aligned}
\end{equation} This is true because the sum of all $r_j$, the count of each letter, is just the length of the multiset.
\end{solution}

\begin{problem}{Easy}
Determine the number of operations involved in each of the components of the procedure to generate a Kronecker graph.
\end{problem}
\begin{solution}
regions: In listing \ref{code:update}, every region is visited and the $update_region$ method is performed. $Update_region$ requires k calculations in the worst case because we must spill over to every index in the region. Thus, our complexity is $O(number_of_regions \cdot k)$.

grass\_hop\_region: From viewing \ref{code:grass_hop}, the number of operations is on the order of the number of edges, $|E|$ times the number of operations in the unrank algorithm. Unranking requires $k^2$ calculations bringing our total count to $O(|E|k^2)$.

mult\_to\_kron: determining the linear index takes $2k$ calculations because we simply have to loop through all the elements in our multiset index and make 2 calculations in the loop. Morton Decoding takes $10k$ calculations because we must loop through 2k digits and do 5 calculations in each iteration. $12k$ calculations is $O(k)$.
\end{solution}


\begin{problem}[Research level with appropriate generality]
Another view of Theorem~\ref{thm:morton} is that we have a permutation induced by
different two different representations of the same number. More specifically, the multi-index representation is in base $n^2$, which we convert to base $n$, and then interleave digits to build a row and column representation. Hence, there are two equivalent representations of the number in base $10$. This hints at a far more general set of permutations based on this type of digit interchange and base-conversion. As an example, the \emph{stride} permutation can be expressed as: represent a number  from $[0,n^2-1]$ in base $n$. This gives a two-digit representation. Swap the digits and convert back to $[0,n^2-1]$. This gives a permutation matrix that exactly corresponds with what is called a stride permutation, which is closely related to the matrix transpose and the difference between row and column major order.  The research problem is to generalize these results. Possible directions include: can any permutation matrix be expressed in this fashion? If not, can you exhibit one and characterize the class of permutations possible?
\end{problem}

\newpage

\bibliographystyle{dgleich-bib}
\bibliography{50-refs}

\section*{Acknowledgements}
We thank Sebastian Moreno for mentioning the original problem during his thesis defense and providing an excellent starting point, as well as Jennifer Neville. Along the way, we received helpful feedback from Eric Chi and C.~Seshadhri. Finally, we wish to thank all of our research team and especially Caitlin Kennedy for a careful read.  This project was supported in part by NSF CAREER CCF-1149756, IIS-1422918, IIS-1546488, and the NSF Center for the Science of Information STC CCF-093937. Gleich's work was also supported by the DARPA SIMPLEX program and the Sloan Foundation.

\end{document}